\documentclass[aps,pre,superscriptaddress, twocolumn, longbibliography]{revtex4-2}
\usepackage{amsmath,graphicx, mathtools,multirow, amssymb}
\usepackage{algorithm, setspace}
\usepackage{algorithmic, bm}
\usepackage[dvipsnames]{xcolor}
\usepackage[colorlinks=true,linkcolor=MidnightBlue,urlcolor=black,
citecolor=MidnightBlue,anchorcolor=MidnightBlue]{hyperref}
\makeatother
\begin{document}
%%==============
%%==============
%%==============
%%==============
\title{Model inference  from fluid flow 
in continuum theories of active matter}
\author{Aditya Mohapatra}
\email{adityamohapatra217@gmail.com} 
\affiliation{Department of Physics, Indian Institute of Technology Madras, Chennai, India}
\affiliation{Institute of Science and Technology Austria (ISTA), Am Campus 1, 3400 Klosterneuburg, Austria}
\author{Sagarika Adhikary} 
\email{a.sagarika@physics.iitm.ac.in} 
\affiliation{Department of Physics, Indian Institute of Technology Madras, Chennai, India} 
\affiliation{School of Physical Sciences, National Institute of Science Education and Research, An OCC of Homi Bhabha National Institute, Jatni-752050, India}
\author{Rajesh Singh} 
\email{rsingh@physics.iitm.ac.in} 
\affiliation{Department of Physics, Indian Institute of Technology Madras, Chennai, India} 
\affiliation{Center for Soft and Biological Matter, IIT Madras, Chennai, India}
\begin{abstract}
    Active matter systems are driven out of 
    thermodynamic equilibrium by localized, microscale energy dissipation. 
    While active field theories
    are highly successful at simulating 
    these non-equilibrium phenomena (the forward problem), characterizing 
    real-world active materials is fundamentally bottle-necked 
    by the difficulty of measuring active stresses directly. 
    This paper addresses the inverse problem using deep learning: 
    model inference and model selection from observable flow field data of active fluids.  
    We formulate an inversion framework  applied
    to two cornerstone paradigms of active continuum physics: Active Model H (representing scalar active matter) and Active Nematics (representing 
    active systems with orientational order). 
    We demonstrate that the kinetic energy spectrum 
    obtained from the fluid flow field preserves a
    high-fidelity signature of activity to infer 
    parameters of active model H and active nematics. 
    The classification of the models requires comparison using conventional numerical work against predicted parameters.
    Our deep learning method presents a principled way to bear upon questions of model inference and selection given the flow field data in continuum models of active matter. 
\end{abstract}
\maketitle
%
%
%
%%====================
%%====================
\section{Introduction}
%%====================
%%====================
%%====================
Active matter defines a fascinating class of non-equilibrium systems composed of individual constituents that consume internal or 
environmental energy to generate spectacular emergent phenomena \cite{cates2011, ramaswamy2010, marchetti2013, te2026colloquium}. 
Continuum theories of active matter 
are written in terms of hydrodynamic variables 
such as density $\rho(\mathbf{r}, t)$, velocity $\mathbf{v}(\mathbf{r}, t)$, and orientation $\mathbf{n}(\mathbf{r}, t)$.
The frameworks of continuum fields successfully capture the large-scale physics of active fluids \cite{marchetti2013, cates2025active, simha2002a, alert2022active, te2026colloquium}.
In traditional theoretical modeling (the forward problem), known activity parameters 
within an active stress tensor are used to predict how the fluid will flow \cite{cates2025active, alert2022active}. Depending on the sign of the parameters, 
the active units act as contractile (puller) or extensile (pusher). 
This forward approach is highly effective for simulating idealized systems. However, when confronting real-world biological systems or complex synthetic microswimmers, a major bottleneck arises: activity is difficult to measure directly.
This work addresses the inverse problem: Can we look at the fluid flow and deduce the underlying activity responsible for the flow?
For active systems at the microscale, the fluid flow operates at a low Reynolds number, 
the surrounding fluid acts as an instantaneous transmitter of 
force densities, governed by the Stokes equation. 
Because the flow field $\mathbf{v}(\mathbf{r})$ is directly coupled to the active stress through the Stokes equation, the fluid flow itself preserves a high-fidelity spatial map of the system's underlying activity. 
Inference of active nematics using machine learning frameworks has received considerable attention recently \cite{colen2021machine,joshi2022, zhou2021, zaplotnik2023neural, piven2024, frishman2021learning, golden2023physically}, while a
systematic inference method of 
scalar field theories of active matter \cite{tiribocchi2015active, singh2019amh} remains unexplored.

In this paper, we present a machine learning framework for inference which can be simultaneously applied to scalar theories of active matter as well as active nematics. By framing the extraction of active properties as a problem of hydrodynamic inference, this paper explores the theoretical and computational frameworks required to reconstruct activity parameters solely from observable flow-field data.  To this end, we utilize 
deep learning and active field theory
to infer 
model parameters and select between models 
using flow field data. 
First, in Sec. \ref{sec:AFT}, we present our theoretical
framework for continuum descriptions of active matter.
This is the forward problem: obtaining fluid flow given activity in distinct models of active matter.
Second, in Sec. \ref{sec:ML}, we show how an ML (machine learning) framework can be deployed to infer  
parameters of active field theory models
from the measured flow field. In particular, we show that the energy spectrum retains a high-fidelity signature of activity that can be used for both model inference and model selection.  
Third, in Sec. \ref{sec:MS}, we present details of the
model selection procedure within our ML framework. 
We note that our model selection procedure requires conventional numerical work with
the ML predicted parameters. Thus, as opposed to parameter inference, the model selection uses some conventional numerical approaches. Developing a fully machine-learning-based model-selection framework remains a promising direction for future work.
Finally, in Sec \ref{sec:summary}, we summarize our results and present future directions. Overall, our results present a principled framework for parameter inference and model selection
from flow field data in continuum models of active matter.  

%
%
%====================================
%====================================
\section{Models of
active fluids}\label{sec:AFT}
%====================================
%====================================
In this section, we present the forward problem 
of computing flow field given a continuum model of active matter.
At the continuum level, the dynamics of active
matter systems is given in terms of active field theories \cite{cates2019active, marchetti2013}.
We consider the two workhorses of
active field theories with momentum conservation.
For systems without any alignment interactions, 
an active field theory
can be given in terms of a scalar order parameter. Such a theory with mass and momentum conservation is referred to as active model H in literature \cite{tiribocchi2015active, singh2019amh}. 
In the presence of alignment interactions, active 
systems can exhibit nematic order. The dynamics of such systems is described using models named as 
active nematics \cite{alert2022active, shankar2022topological, ramaswamy2017active, marchetti2013}. We describe both these models of active matter in detail below.

\subsection{Active model H} \label{sec:amh}
Active Model H  is used 
to describe the hydrodynamic behavior of scalar active matter, 
such as self-propelled microswimmers 
suspended in a 
momentum-conserving solvent without explicit orientational alignment. 
The nomenclature of active model H (AMH) follows from the Hohenberg-Halperin classification \cite{hohenberg1977theory}. 
While the traditional, passive Model H couples the diffusion of a conserved scalar field (like particle concentration) to a fluid velocity field, its active counterpart injects non-equilibrium physics by introducing an intrinsic active stress tensor that breaks time-reversal symmetry. For contractile active particles, this microscale driving force functions effectively as a negative interfacial tension, causing fluid flow to spontaneously stretch phase interfaces \cite{tiribocchi2015active, singh2019amh}.
Spontaneous stretching of interfaces for contractile active stress
in active model H
leads to a self-shearing instability \cite{singh2019amh}. 
The result is
a dynamical steady-state of active fluids 
maintained by the competing Ostwald ripening and 
a self-shearing instability. 
The dynamical steady-state of active model H for contractile activity
has also been 
referred to 
as scalar active turbulence in the literature \cite{padhan2025cahn, radhakrishnan2026irreversibility}.

% \subsubsection{Equation of Motion}
We now present the forward problem to describe the dynamics of
active model H for a prescribed activity. 
Consider a conserved scalar field $\phi(\mathbf{r},t)$
in a momentum-conserving fluid with velocity field $\mathbf{v}(\mathbf{r},t)$. 
The dynamics of the scalar field is then given in terms of contributions to the flux from the mechanical sector ($\mathbf{v}\phi$) and the diffusive sector as:
\begin{subequations}
\label{eq:AMH}
\begin{align}
{\partial_t \phi} 
&=
-\bm{\nabla}\cdot
\left (%\mathbf{J}
\phi \mathbf v
-\boldsymbol{\nabla}\frac{\delta\mathcal{F}}{\delta\phi}
\right)
,\\
\mathcal{F}[\phi]&=\int\left(-\frac{{b}}{2}\phi^{2}+\frac{b}{4}\phi^{4}+\frac{\kappa_p}{2}(\mathbf{\nabla}\phi)^{2}\right)d\mathbf{r}.
\label{eq:orderPara}%\\
% \mathbf{J}=\left(-\boldsymbol{\nabla}\frac{\delta\mathcal{F}}{\delta\phi}
% % \mu+\zeta(\nabla^{2}\phi)\boldsymbol{\nabla}\phi
% \right).
%+\sqrt{2DM}\boldsymbol{\Lambda_\phi}.
%,\\
% \mu=\mu^{E}+\mu^{\lambda},\quad\mu^{E}=\frac{\delta\mathcal{F}}{\delta\phi},\quad\mu^{\lambda}=\lambda|\boldsymbol{\nabla}\phi|^{2}.
\end{align}
\end{subequations}
% Here, $T$ is the temperature.
% The noise term $\boldsymbol{\Lambda}_{\phi}$ is
% Gaussian with zero mean and correlations given by 
% \begin{equation}
% \big\langle\Lambda_{\phi,\alpha}({\bf r},t)\Lambda_{\phi,\beta}({\bf r}',t')\big\rangle=\delta_{\alpha\beta}\delta({\bf r}-{\bf r}')\delta(t-t').\label{eq:lambda_phi}
% \end{equation}
Here, ${\cal F}$ is the Landau-Ginzburg free energy functional, while $b$ and $\kappa_p$ are parameters of the free energy. Both these constants are positive: $b>0$ and $\kappa_p>0$.
% \[
% \]
%%================
%%================
\begin{figure}[t]
    \centering
    % \linewidth will now automatically span both columns
    % \includegraphics[width=0.99\textwidth]{amh_an.jpg}
    \includegraphics[width=0.44\textwidth]{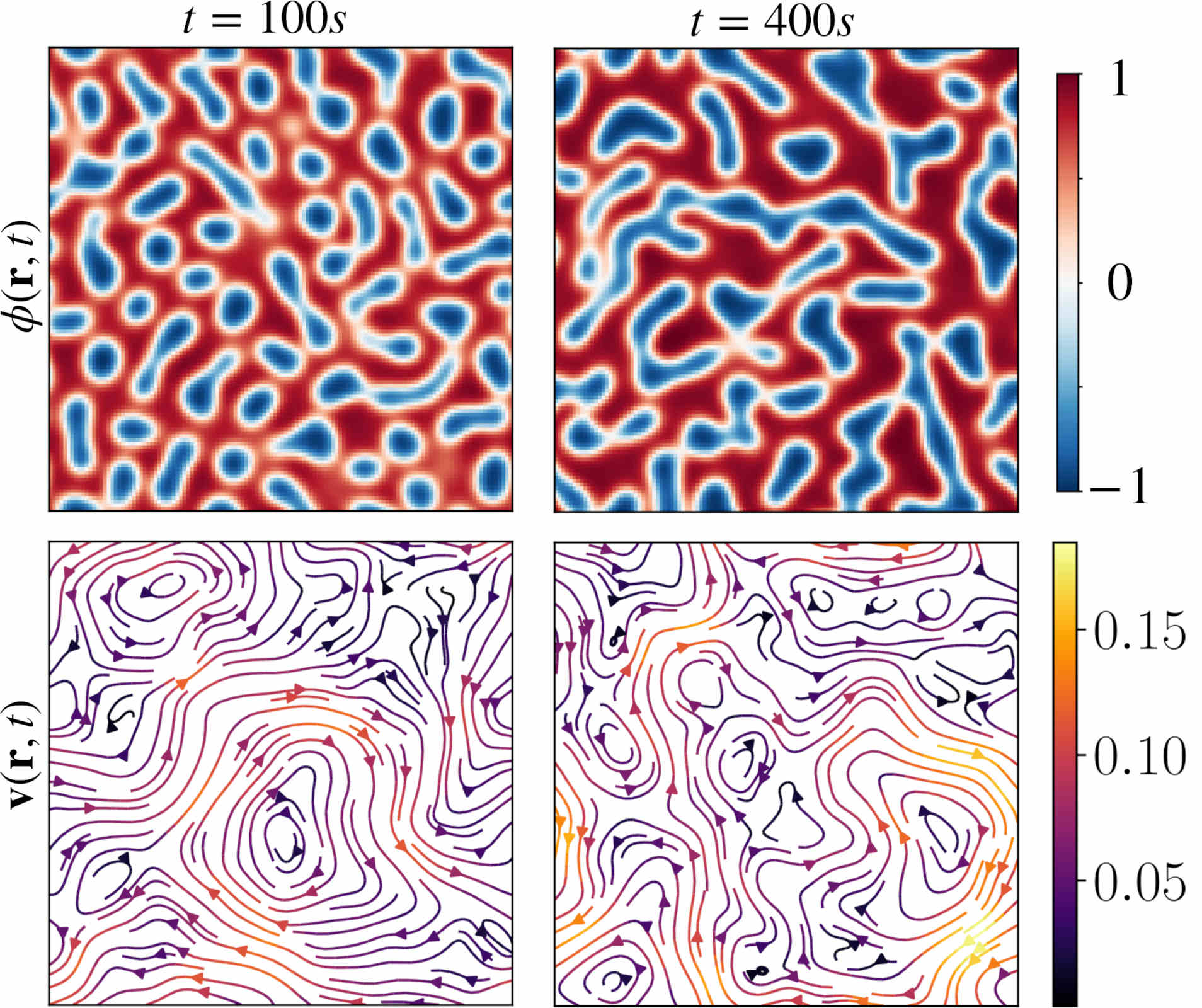}    
    \caption{\textbf{Active Model H}. 
    %\textbf{Snapshots and energy spectrum of active model H and active nematics}. 
   Snapshots from simulation of Active Model H 
    for the order-parameter field $\phi(\mathbf{r},t)$ (top row)
and the induced velocity field $\mathbf{v}(\mathbf{r},t)$ (bottom row)
at $t = 100$s,  and $400$s
for $\kappa_p = 1.0$, $\tilde{\kappa} = -0.5$, $b = 0.5$.
Colour in the top row maps $\phi \in [-1,1]$;
streamlines in the bottom row are coloured by the local speed
$|\mathbf{v}|$. 
}
    \label{fig:amh}
\end{figure}
%%================
The fluid velocity $\mathbf v$, in the limit of low Reynolds number (as applicable
to microswimmers), is obtained from the solution of the Stokes equation:
\begin{equation}
\bm{\nabla}\cdot\bm{\sigma} 
=-\bm{\nabla}\cdot(\bm{\Sigma}^{\mathrm{A}}
+\bm{\Sigma}^{\mathrm{E}}).
\label{eq:stokes}
\end{equation}
Here $\bm{\sigma} $ is the Cauchy stress tensor, which is given as:
\begin{align}
  % {\sigma}_{ij}=-p\delta_{ij}+\eta( 
  % {\nabla}_i  {v}_j+
  % {\nabla}_j  {v}_i
  % )
\bm{\sigma}=-p\mathbf{I}+\eta
\left[
\bm{\nabla}\mathbf{v}+(\bm{\nabla}\mathbf{v})^{T}
\right]
\label{eq:CStress}
\end{align}
$\eta$ is viscosity, $\mathbf{I}$ is
the identity tensor, and $p$ is the pressure field which contains
all isotropic terms and ensures incompressibility ($\nabla\cdot\mathbf{v}=0$)
\citep{landau1959fluid}. The deviatoric stresses $\mathbf{\Sigma}^{\mathrm{E}}$
and $\mathbf{\Sigma}^{\mathrm{A}}$ are then, in $d$-dimensions, given to
the required order as: 
\begin{gather}
\mathbf{\Sigma}^{\mathrm{E}}=-\kappa_p\mathbf{S},\qquad
\mathbf{\Sigma}^{\mathrm{A}}=-(\tilde{\kappa}-\kappa_p)\mathbf{S}.
\\
\label{eq:activeStress}
    \mathbf{S}\equiv(\mathbf{\nabla}\phi)(\mathbf{\nabla}\phi)-\tfrac{1}{d}|\mathbf{\nabla}\phi|^{2}\mathbf{I}.
\end{gather}
It is worthwhile to note that 
the mechanical stress in the above equations 
$\boldsymbol{\Sigma}^{A}+\boldsymbol{\Sigma}^{E}=-\tilde{\kappa}\boldsymbol{S}$ 
is not derived from a free energy. 
Thus, in general, it breaks detailed balance. 
The coefficient $\tilde\kappa$ 
can be either positive (for extensile microswimmers) or negative (for contractile microswimmers) 
\cite{tiribocchi2015active, singh2019amh, cates2025active} 
unlike equilibrium systems where $\kappa_p>0$ and the limit $\boldsymbol{\Sigma}^{A}=0$ holds. In Appendix \ref{sec:appMethod}, we explain our numerical method to solve the above equations.
In
Fig.\ref{fig:amh}, we show two snapshots from the steady state of active model H for the order parameter as well as the corresponding velocity field.
\textcolor{black}{
We demonstrate robustness of the inference framework by 
studying the role of an additional term, $\lambda|\nabla\phi|^2$  \cite{wittkowski2014scalar}, to Eq.\eqref{eq:AMH} in Appendix \ref{app:lambda}. }

%%================
\begin{figure}[t]
    \centering
    % \linewidth will now automatically span both columns
    % \includegraphics[width=0.99\textwidth]{amh_an.jpg}
    \includegraphics[width=0.44\textwidth]{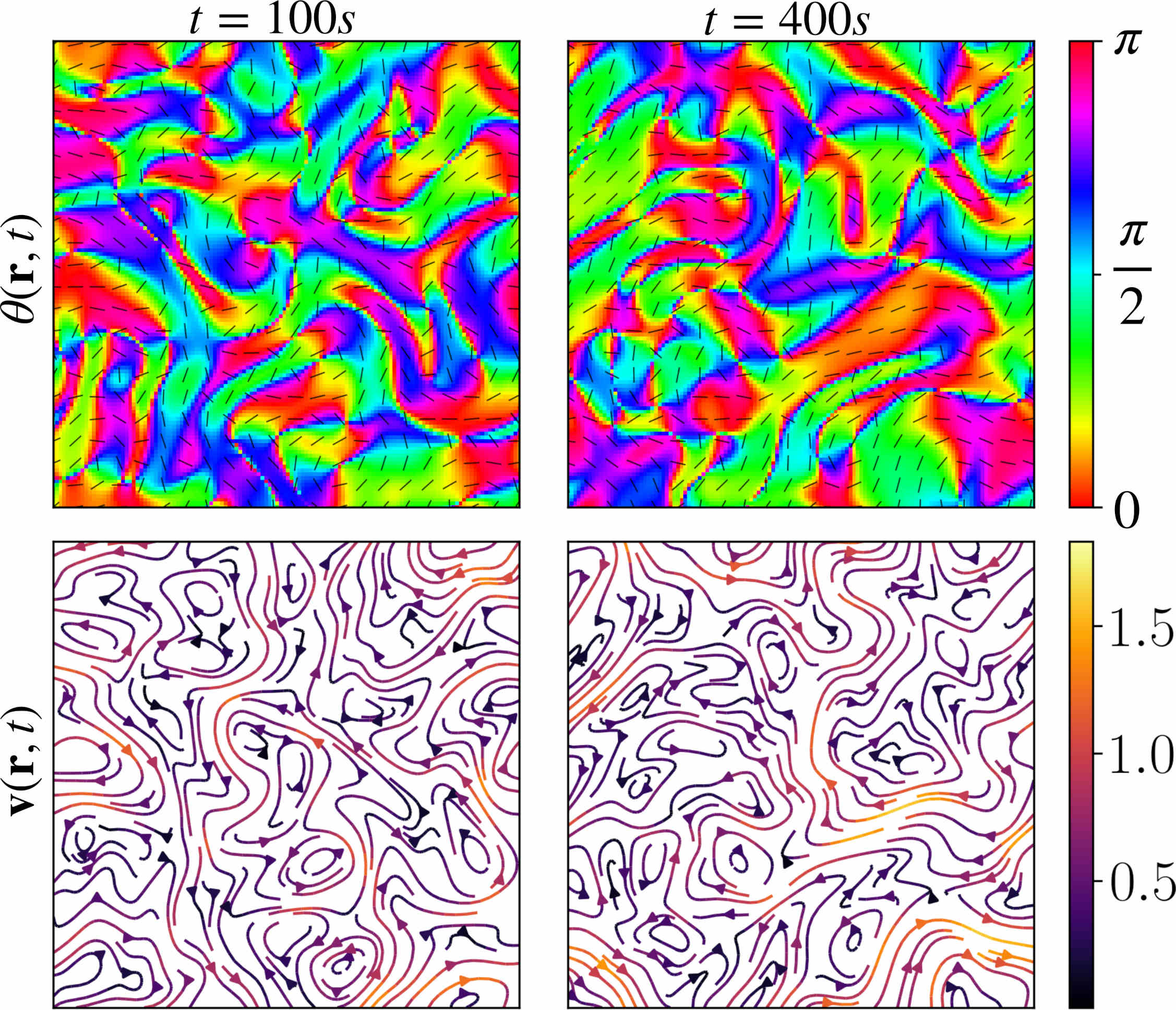}    
    \caption{\textbf{Active Nematics}. 
    %\textbf{Snapshots and energy spectrum of active model H and active nematics}. 
Snapshots of the nematic director field $\theta(\mathbf{r},t)$ (top row)
and the induced velocity field $\mathbf{v}(\mathbf{r},t)$ (bottom row)
at two time instances ($t = 100$s, and $400$s)
for $K = 1.0$, $\zeta = -0.5$.
Colour in the top row maps the director field $\theta \in [0,\pi]$;
streamlines in the bottom row are colored by the local speed
$|\mathbf{v}|$.
}
    \label{fig:an}
\end{figure}
%%================
\subsection{Minimal Description of Active Nematics} \label{sec:an}
Active Nematics \cite{marchetti2013, ramaswamy2017active, doostmohammadi2018active, alert2020universal, mukherjee2023intermittency, alert2022active, simha2002a} extend the classical continuum hydrodynamics of passive nematic liquid crystals frameworks 
to systems of elongated, self-propelled units that exhibit long-range orientational alignment. While passive nematics relax toward uniform alignment to minimize elastic free energy, active nematics are driven far from equilibrium by localized, dipolar forces generated along the long axis of each constituent, such as crawling spindle-shaped cells or kinesin-driven microtubule bundles. This intrinsic active stress renders the uniformly aligned state inherently unstable, triggering a hydrodynamic
instability that continuously disrupts the orientational order. Active nematic systems are modeled within a continuum hydrodynamic framework in which the coarse-grained velocity field $\mathbf{v}(\mathbf{r},t)$ is
coupled to the nematic order parameter tensor $Q_{ij}(\mathbf{r},t)$ \cite{ramaswamy2017active, marchetti2013}. 

As described in the previous section of active model H, the fluid flow is given by the Stokes equation.
The incompressibility condition enforces
$\nabla \cdot \mathbf{v} = 0$.
For the case of active nematics,
the fluid flow is driven by active stress ${\Sigma}_{ij}^{\mathrm{AN}}$:
\begin{equation}
 {\nabla}_i {\sigma}_{ij} +
 {\nabla}_i
 {\Sigma}_{ij}^{\mathrm{AN}} =0\label{eq:stokesAN}
\end{equation}
Here, the Roman alphabets $i,j$ denote Cartesian indices of a tensor. 
The Cauchy stress $\sigma_{ij}$ was defined in Eq.\eqref{eq:CStress}.
The active stress $ {\Sigma}_{ij}^{\mathrm{AN}}$ for AN (active nematics)  is:
\begin{align}
% \sigma_{ij} &= -p \delta_{ij} + 2\eta E_{ij},
 {\Sigma}_{ij}^{\mathrm{AN}}&=
- \zeta Q_{ij},
\label{eq:activeStressAN}
\end{align}
where $\zeta$ is the activity parameter (positive for extensile and negative
for contractile systems). In this paper, we only consider contractile systems for simplicity.
The nematic order parameter
is expressed in terms of the director field $\mathbf{n} = (\cos\theta,\sin\theta)$ as
\begin{align}
Q_{ij} = \left(n_i n_j - \tfrac{1}{d}\delta_{ij}\right).
\end{align}
Here, $d$ is the spatial dimension. For two-dimensional systems ($d=2$),
a minimal description of active nematics can be formulated in terms of the
coarse-grained director angle field $\theta(\mathbf{r},t)$. The dynamics
of $\theta$ is governed by \cite{alert2020universal}:
\begin{align}
\partial_t \theta 
=
-\bm{\nabla}\cdot
\left (%\mathbf{J}
\theta \mathbf v
\right)
+  K \nabla^2 \theta +   (\partial_x v_y - \partial_y v_x)
\label{eq:theta}
\end{align}
Here, $K$ is the Frank elastic constant, while the Frank free energy is \cite{julicher2018hydrodynamic, alert2020universal}:
\begin{align}
        F^n =\frac{K}2 \int |\bm \nabla \theta|^2 d\mathbf r
\end{align}
Together,  Eq.\eqref{eq:activeStressAN} and Eq.\eqref{eq:theta} present a minimal hydrodynamic theory of active nematic fluids \cite{alert2020universal}. 
We have used these equations to generate data for inference framework described below. In
Fig.\ref{fig:an}, we show two snapshots from the steady state of active nematic for the order parameter as well as the corresponding velocity field.

%%================
\begin{figure}[b]
    \centering
    % \linewidth will now automatically span both columns
    % \includegraphics[width=0.99\textwidth]{amh_an.jpg}
    \includegraphics[width=0.46\textwidth]{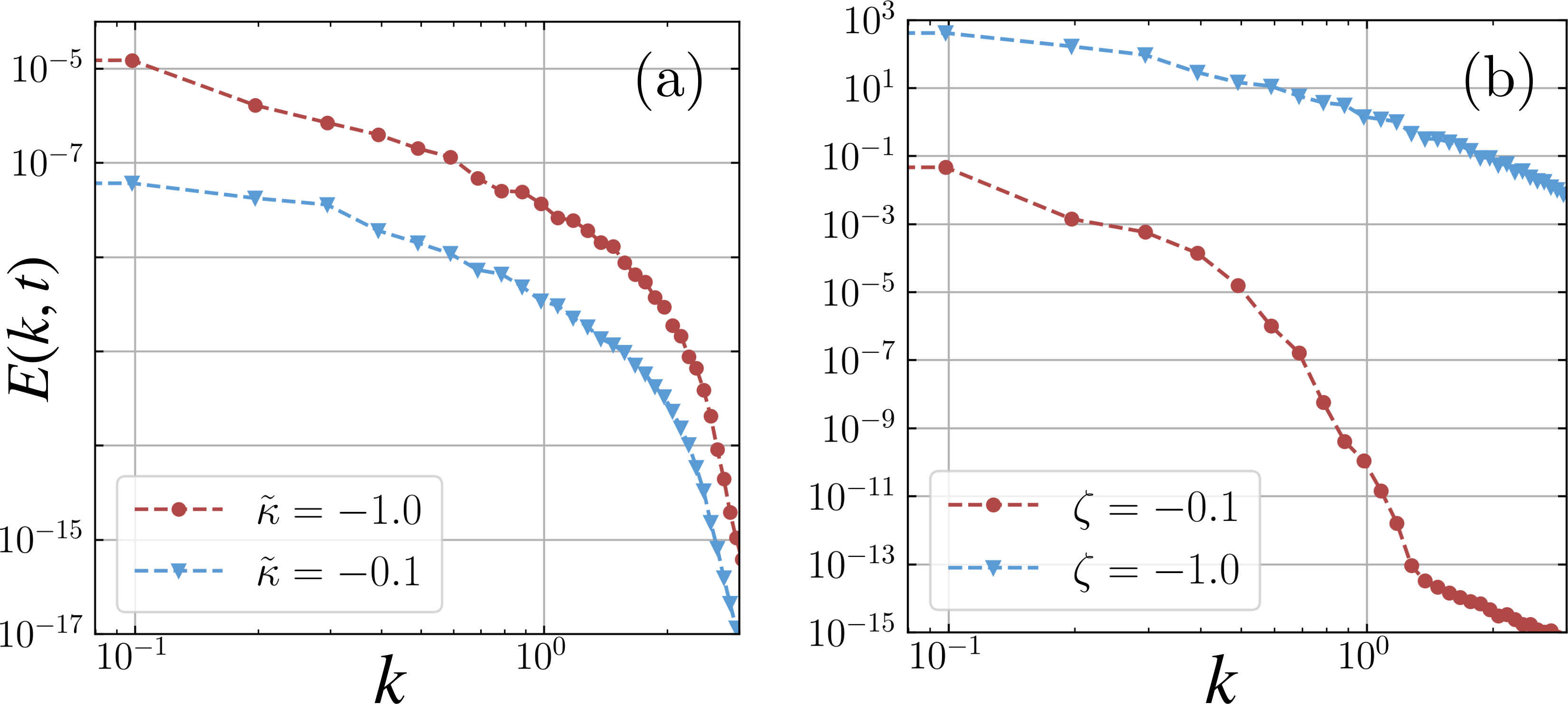}    
    \caption{\textbf{Kinetic Energy spectrum}. 
Panel (a)
 is the energy Spectrum $E(k,t)$  of active model H
    for two different values of activity $\tilde \kappa=-0.1$ and $-1.0$, while panel (b)
 is energy spectrum $E(k,t)$  of active nematics
    for two different values of activity $\zeta=-0.1$ and $-1.0$.
% [SPECTRUM PLOTS TO CHANGE]
}
    \label{fig:amh_an_ek}
\end{figure}
%%================

%%----------------------------------
%%----------------------------------
\subsection{Kinetic Energy spectrum}
%%----------------------------------
%%----------------------------------
To develop the machine learning framework, as we describe below, 
we analyze
the spectra of the kinetic energy $E(k)$. The shell-averaged
kinetic energy spectrum $E(k)$ is defined as
\begin{align}
E(k,t) &\equiv \frac{1}{n(k)}\sum_{k \le k' < k+\Delta k} \frac{1}{2}\big|\hat{\mathbf{v}}(\boldsymbol{k}',t)\big|^2
\end{align}
Here $\hat{\mathbf{v}}(\boldsymbol{k}',t)$ is the discrete Fourier transform of
the velocity field $\mathbf{v}$, and $n(k)$ is the number of discrete
wave-vectors $\boldsymbol{k}'$ satisfying $k \le |\boldsymbol{k}'| < k+\Delta k$. See
appendix \ref{sec:appMethod} for details. The summation condition below the
sigma symbol: $k \le k' < k+\Delta k$ means taking a spherical shell of radius $k$
and thickness $\Delta k$, and the pre-factor $1/n(k)$ averages the kinetic
energy over every wavenumber vector whose magnitude $k' = |\boldsymbol{k}'|$
falls inside that specific shell.
Energy spectrum is known to have distinct signatures of activity in continuum models of active matter \cite{padhan2025cahn, radhakrishnan2026irreversibility, alert2020universal}. We use this idea to present a novel inference framework using deep learning. 

% %%====================
% \subsection{Enstrophy spectrum}
% %====================
% Following the analysis of the previos section, we also define the spectrum of enstrophy in this section. 
% \begin{align}
%     \mathcal{E} = \int \omega^2 d\mathbf r
%     \label{eq:ENS}
% \end{align}
% Here,  
% $\boldsymbol{\omega}=\tfrac{1}{2}\boldsymbol{\nabla}\times\mathbf{v}$
% is the vorticity.
% We define the spectrum of the enstrophy,
% $\mathcal{E}(k)$, using Fourier transform. See appendix \ref{sec:appMethod} for our definition. We have:
% \begin{align}
%     \frac{\langle \mathcal{E}\rangle}{L^2} = \int \mathcal{E}(k) dk
% \end{align}
% Using the definition of kinetic energy in Eq\eqref{eq:ENS}, we obtain:
% \begin{align}
%     \mathcal{E}(k)=\pi L^{2}k|\omega_{k}|^{2}
% \end{align}

We plot the kinetic energy spectrum $E(k,t)$
  in Fig.\ref{fig:amh_an_ek}(a), and (b) for  active model H and active nematics respectively. It can be seen that it is sensitive to the choice of parameters and model. We also find that it changes as a function of time (not shown here). In the following, we use the kinetic energy spectrum to develop our ML model to infer parameters. Usually, in the literature, the energy spectrum $E(k)$ is averaged over time. We use the temporal information in the spectrum to infer all the parameters and select between models. 

\begin{figure}[t]
    \centering
    \includegraphics[width=0.32\textwidth]{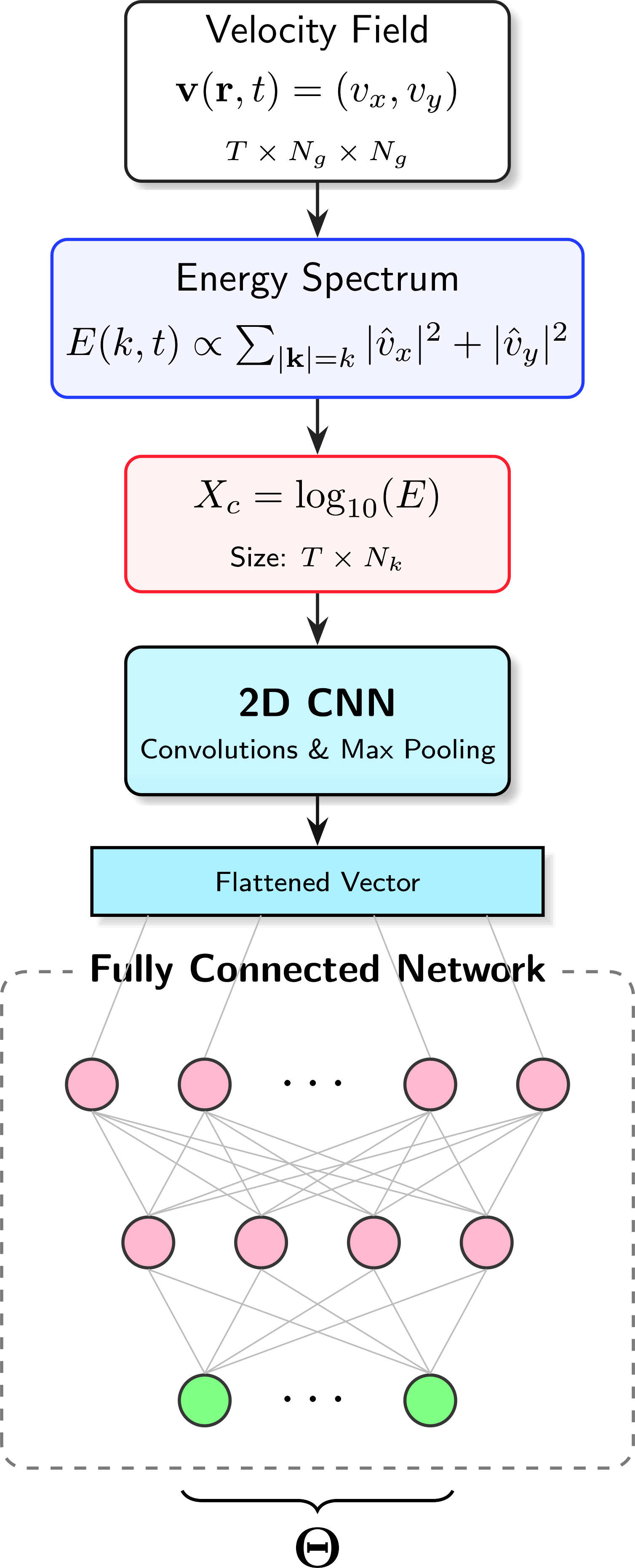} 
    \caption{\textbf{Machine Learning Framework}. The velocity field is used to compute the kinetic energy spectrum $E(k,t)$.
    The spectrum is log-transformed and passed through a 2D CNN, whose output is flattened and mapped through fully connected layers to predict the model parameters $\Theta$: $(\tilde\kappa, b, \kappa_p)$ for Active Model H and $(\zeta, K)$ for Active Nematics.}
\label{fig:network_architecture}
\end{figure}

%%%=================================
%%%=================================
%%%=================================
\section{Machine learning framework}
\label{sec:ML}
%%%=================================
%%%=================================
%%%=================================
We formulate parameter inference as a supervised regression problem in which the objective is to predict the governing model parameters from the kinetic energy spectrum, $E(k,t)$, computed from flow fields. To this end, we train two convolutional neural networks (CNNs)\cite{lecun1998gradient, goodfellow2016deep}, $\mathcal{N}_{\mathrm{AMH}}$ and $\mathcal{N}_{\mathrm{AN}}$, corresponding to Active Model H and Active Nematics, respectively. Although both networks operate on the same input representation, each is trained to predict the parameter set specific to its underlying continuum model. Specifically, $\mathcal{N}_{\mathrm{AMH}}$ predicts the parameters $(\tilde{\kappa}, b, \kappa_p)$, while $\mathcal{N}_{\mathrm{AN}}$ predicts $(\zeta, K)$.

For each trajectory, the kinetic energy spectrum, $E(k,t)$, is evaluated over the late-time statistically stationary regime and represented as a two-dimensional array with wavenumber and time as its axes. To improve numerical stability and compress the large dynamic range of the spectral values, the network input is defined as the logarithmically transformed spectrum,
$X_c = \log_{10}\left(E(k,t) + \varepsilon\right)$
where $\varepsilon$ is a small positive constant introduced to avoid numerical singularities.

\subsection{Network architecture}

The transformed spectrum, $X_c$, is treated as a single-channel image and processed by a convolutional neural network (Fig.~\ref{fig:cnn_architecture} of Appendix), exploiting local correlations along both the spectral and temporal axes in a manner analogous to spectrogram-based classification \cite{lecun1998gradient, hershey2017cnn}.
The network consists of two convolutional blocks followed by a fully connected regression head. The convolutional blocks employ $5\times5$ and $3\times3$
kernels, respectively, with LeakyReLU\cite{maas2013rectifier} activations and $2\times2$ max-pooling after each convolution. \textcolor{black}{The shared convolutional filters provide an efficient way to detect local patterns, such as peaks and slopes, across neighboring wavenumber bins and time steps. This local feature extraction motivates using a CNN rather than a purely fully connected network, which does not explicitly exploit the local structure of the spectrum.} The extracted features are flattened and mapped to the
model parameters through two fully connected hidden layers and a final linear output layer (Fig.~\ref{fig:network_architecture}). \textcolor{black}{The fully connected layers preserve information about feature locations, so
the network does not impose translation invariance along the wavenumber or
time axes, consistent with the symmetries of the system.} The two networks differ
only in the dimension of the output layer.

%%%--------------
%%%--------------
\begin{figure*}[t]
    \centering
    \includegraphics[width=0.8\textwidth]{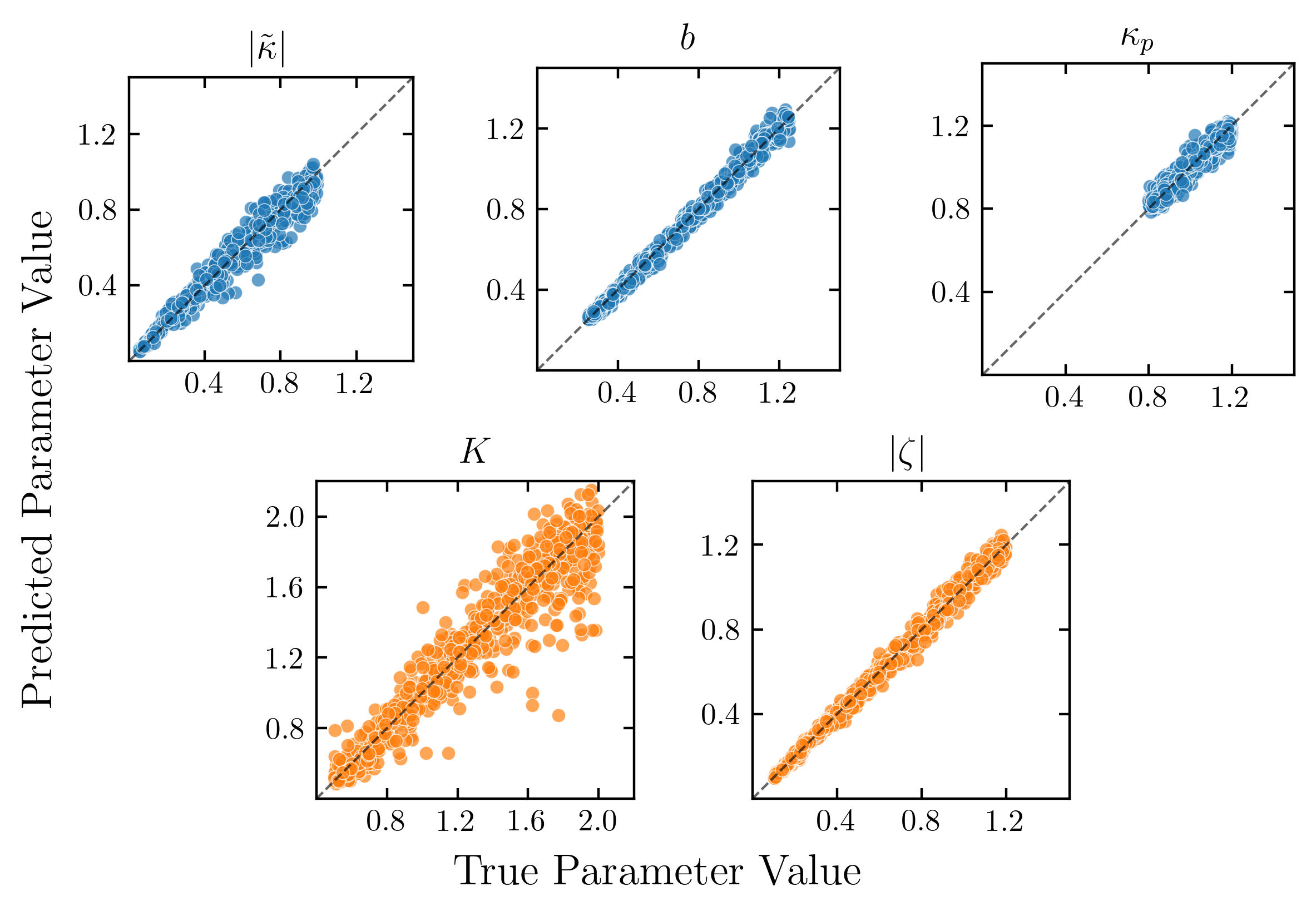}
    \caption{\textbf{Parameter recovery.} Predicted versus true parameter
    values on the held-out test set, for Active Model H (top row, green:
    $\tilde{\kappa}$, $b$, $\kappa_p$) and Active Nematics (bottom row,
    orange: $K$, $\zeta$). Each point is one test simulation; the dashed
    line marks perfect recovery ($y=x$). }
    \label{fig:param_recovery}
\end{figure*}
%%============
%%%--------------

\subsection{Training}

Separate training datasets are generated for each continuum model by sampling the physically relevant parameter space while excluding the vicinity of the passive limit, where the activity is too weak to produce distinctive flow patterns. Parameter values are sampled using a combination of a jittered Cartesian grid and random sampling to ensure both broad coverage of the parameter space and dense sampling around representative parameter combinations.

Since the activity parameters, $\tilde{\kappa}$ for Active Model H and $\zeta$ for Active Nematics, have fixed signs by convention (corresponding to contractile or extensile activity), the networks are trained to predict the logarithm of their magnitudes. The appropriate sign is then restored after inference according to the prescribed convention. The logarithmic transformed targets are standardized to zero mean and unit variance using a scaler fit on the training set; this scaler is saved and reapplied to invert the predictions back to physical units at the test time.

Networks are implemented in PyTorch \cite{paszke2019pytorch} and optimized using the AdamW \cite{loshchilov2019decoupled} optimizer with a mean-squared-error loss and an 85/15 train-validation split. Model performance is assessed using an independent test set comprising simulations not encountered during training. Prediction accuracy is quantified using the mean absolute percentage error (MAPE) between the inferred and true parameter values. The training and validation loss curves, along with the per-parameter percentage error, are shown in Fig.~\ref{convergence_amh} of Appendix~\ref{sec:appMethod}.

\subsection{Parameter recovery}

%%%--------------

To assess the accuracy of the trained networks, we evaluate parameter
recovery on an independent test set of simulations not used during
training or validation. Figure~\ref{fig:param_recovery} shows the
predicted versus true values for all inferred parameters:
$(\tilde{\kappa}, b, \kappa_p)$ for Active Model H and $(\zeta, K)$ for
Active Nematics. 
% Predictions for $b$, $K$, and $\zeta$ lie close to the
diagonal across the sampled range. 
Overall, the tight clustering along $y=x$ demonstrates that the kinetic energy spectrum retains sufficient information to recover the activity parameters directly from flow-field data, without any reference to the underlying order-parameter or nematic director fields. This is consistent with the stable convergence of training and validation loss shown in Fig.~\ref{convergence_amh} of Appendix.

Since the energy spectrum carries distinct signatures of activity in these continuum models, as noted in Sec.~\ref{sec:AFT}, the map from activity parameters to $E(k,t)$ is expected to be effectively one-to-one over the sampled range. Working with the time-resolved spectrum, rather than its time average, further resolves parameter combinations that converge to similar steady states but differ in fluctuations in time. The absence of deviation in parameter recovery (Fig.~\ref{fig:param_recovery}) and of a validation-loss plateau during training (Fig.~\ref{convergence_amh} of  the appendix) is consistent with this map being locally invertible over the sampled parameter range.

{\color{black}
To report the uncertainty in a parameter estimate, we construct an empirical
$90\%$ interval from the parameter-recovery errors on 1400 held-out simulations.
The interval details and their coverage on held-out AMH and AN test cases are
reported in Appendix~\ref{app:param_uncertainty},
Table~\ref{tab:param_uncertainty}. These intervals are calibrated for data
generated under the tested conditions; they are not intrinsic uncertainty
estimates produced by the networks.
}

%%================
\begin{figure}[t]
    \centering
    \includegraphics[width=0.46\textwidth]{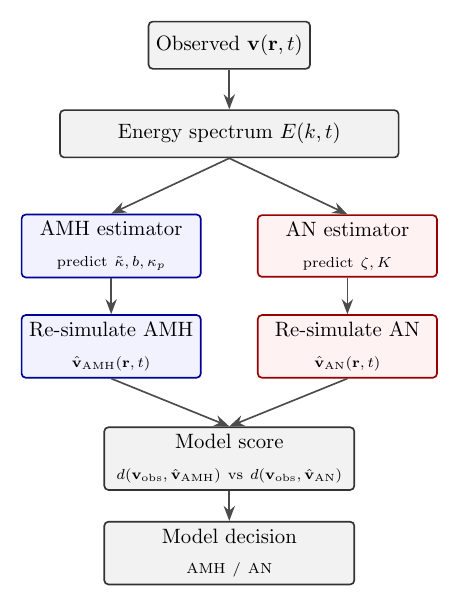}
    \caption{\textbf{Model selection flowchart.}
Given a flow field data, we first compute the parameters of the model. This is then used to run simulation and compute spectrum. Time average spectrum is compared with the data provided to do model selection. \textcolor{black}{
Our model selection procedure allows to choose neither of the two (NOTT)  in case of not good fit to either to the two candidate models (AMH or AN). 
}
A detailed algorithm is given in Algorithm \ref{alg:modelS}.
}
\label{fig:model_selection}
%%================
    % (A) Normalized confusion matrix for the AMH/AN decision on held-out
    % test simulations, reported as the per-true-class accuracy (with raw
    % counts in parentheses) so that any class-imbalanced errors are
    % visible directly.
    % (B) Kernel density estimate of the log discrepancy ratio
    % $\log_{10}(d_{\mathrm{AN}}/d_{\mathrm{AMH}})$, split by the true
    % generating model; the dashed line marks the decision boundary at
    % zero, and the clean separation of the two distributions around it
    % reflects the discriminating power of the resimulation step.
    % (C) Discrepancy scores $d_{\mathrm{AN}}$ versus $d_{\mathrm{AMH}}$
    % for each test sample on log-log axes, with the equal-score diagonal
    % shown for reference; true-AMH samples (green circles) fall
    % consistently below the diagonal and true-AN samples (orange squares)
    % consistently above it.
    % (D) Distribution of the assigned confidence $P(M^*)$, split by true
    % model, with individual samples colored by whether the selection was
    % correct (green) or incorrect (red).}
\end{figure}
%%================
% %================
% \begin{figure}[b]
%     \centering
%     % Adjust the 0.9\linewidth if you need it to be slightly smaller or larger across the two columns
%     \includegraphics[width=0.82\columnwidth]{modeS_pipeline.jpg} 
%     \caption{\textbf{Model Selection pipeline}.
%     }
%     \label{fig:model_selection}
% \end{figure}
% %%=======================

%%
%%

%%================
%%
%========================
%========================
%========================
\section{Model Selection}
\label{sec:MS}
%========================
%========================
%========================
Beyond estimating parameters within a single model, 
we can also predict
which model (AMH or AN) better suits a given velocity field data \cite{mackay2003information, sivia2006data}. 
A trained
network will output parameter values regardless of whether its own model
actually describes the observed data, so parameter estimation by itself
cannot tell us which model is the right one \cite{jaynes2003probability}. To decide between AMH and AN,
we go one step further and check each candidate model against the data by
re-simulating it \cite{cranmer2020frontier, toni2008approximate, singh2018fast, sisson2018handbook}, following the procedure in Algorithm~\ref{alg:modelS} 
and
Fig.~\ref{fig:model_selection}.
%

%========================
% \begin{figure}[t]
% \begin{algorithm}[H]
% \begin{spacing}{1.2}
% \caption{Model Selection}
% \label{alg:modelS}
% \begin{algorithmic}[1]
% \vspace{0.25cm}
% \REQUIRE $v_x(\mathbf{r},t), v_y(\mathbf{r},t)$; $\mathcal{N}_{\mathrm{AMH}}, \mathcal{N}_\mathrm{AN}$; $R$
% \ENSURE $M^*, \theta^*$
% \STATE $\mathcal{M} \gets \{\mathrm{AMH}, \mathrm{AN}\}$
% \STATE Compute $X_c \gets \log_{10}(E(k,t)+\varepsilon)$ from the observed fields
% \STATE $\theta_{\mathrm{AMH}} = (\tilde\kappa, b, \kappa_p) \gets \mathcal{N}_{\mathrm{AMH}}(X_c)$
% \STATE $\theta_\mathrm{AN} = (\zeta, K) \gets \mathcal{N}_\mathrm{AN}(X_c)$
% \STATE Compute the time-averaged spectrum $\log\langle E\rangle_t(k)$ of the observed fields
% \FOR{$c \in \mathcal{M}$}
%   \FOR{$r=1$ to $R$}
%     \STATE $(\hat{v}_x^{(r)},\hat{v}_y^{(r)}) \gets \mathcal{S}(\theta_c)$ \COMMENT{new random initial condition}
%     \STATE Compute $\log\langle \hat{E}^{(r)}\rangle_t $
%   \ENDFOR
%   \STATE $d_c \gets \dfrac{1}{R N_k} \sum_{r,k} \big[\log\langle E\rangle_t  - \log\langle \hat{E}^{(r)}\rangle_t \big]^2$
% \ENDFOR
% \STATE $M^* \gets \arg\min_{c} d_c$, \quad $\theta^* \gets \theta_{M^*}$
% \RETURN $M^*, \theta^*$
% \end{algorithmic}
% \end{spacing}
% \end{algorithm}
% \end{figure}

\begin{figure}[t]
\begin{algorithm}[H]
\begin{spacing}{1.2}
\caption{\textbf{Model Selection algorithm}. 
\textcolor{black}{}
\label{alg:modelS}
}
\begin{algorithmic}[1]
\vspace{0.25cm}
\REQUIRE $v_x(\mathbf{r},t),v_y(\mathbf{r},t)$, $\mathcal{N}_{\mathrm{AMH}},\mathcal{N}_{\mathrm{AN}}$, $R$,\textcolor{black}{$ a,\{T_c\}$}
\ENSURE \textcolor{black}{$M^*, \theta^*, P_{\mathrm{AMH}}, P_{\mathrm{AN}}$}
\STATE $\mathcal{M} \gets \{\mathrm{AMH}, \mathrm{AN}\}$
\STATE Compute $X_c \gets \log_{10}(E(k,t)+\varepsilon)$ from the observed fields
\STATE $\theta_{\mathrm{AMH}} = (\tilde\kappa, b, \kappa_p) \gets \mathcal{N}_{\mathrm{AMH}}(X_c)$
\STATE $\theta_\mathrm{AN} = (\zeta, K) \gets \mathcal{N}_\mathrm{AN}(X_c)$

\STATE Compute the time-averaged spectrum $\log\langle E\rangle_t(k)$ of the observed fields
\FOR{$c \in \mathcal{M}$}
  \FOR{$r=1$ to $R$}
    \STATE $(\hat{v}_x^{(r)},\hat{v}_y^{(r)}) \gets \mathcal{S}(\theta_c)$ \COMMENT{new random initial condition}
    \STATE Compute $\log\langle \hat{E}^{(r)}\rangle_t $
  \ENDFOR
  \STATE $d_c \gets \dfrac{1}{R N_k} \sum_{r,k} \big[\log\langle E\rangle_t  - \log\langle \hat{E}^{(r)}\rangle_t \big]^2$
  \STATE \textcolor{black}{$G_c\gets d_c/T_c$}
\ENDFOR
\STATE $M^* \gets \arg\min_{c} d_c$, \quad $\theta^* \gets \theta_{M^*}$

% \RETURN $M^*, \theta^*$

\STATE \textcolor{black}{$\ell\gets\log_{10}(d_{\mathrm{AN}}/d_{\mathrm{AMH}})$}
\STATE \textcolor{black}{$P_{\mathrm{AMH}}\gets[1+\exp(-a\ell)]^{-1}$}
\STATE \textcolor{black}{$P_{\mathrm{AN}}\gets1-P_{\mathrm{AMH}}$}

\STATE \textcolor{black}{$\mathcal A\gets\{c\in\mathcal M:G_c\leq1\}$}
\IF{$\mathcal A=\varnothing$}
    \STATE \textcolor{black}{$M^*\gets\texttt{NOTT}$}
\ELSE
    \STATE \textcolor{black}{$M^*\gets\arg\min_{c\in\mathcal A}d_c$}
    \STATE \textcolor{black}{$\theta^*\gets\theta_{M^*}$}
\ENDIF
\RETURN \textcolor{black}{$M^*,\theta^*,P_{\mathrm{AMH}},P_{\mathrm{AN}}$}
\end{algorithmic}
\end{spacing}
\end{algorithm}
\end{figure}
%========================

%%================
\begin{figure*}[t]
    \centering
    \includegraphics[width=0.88\textwidth]{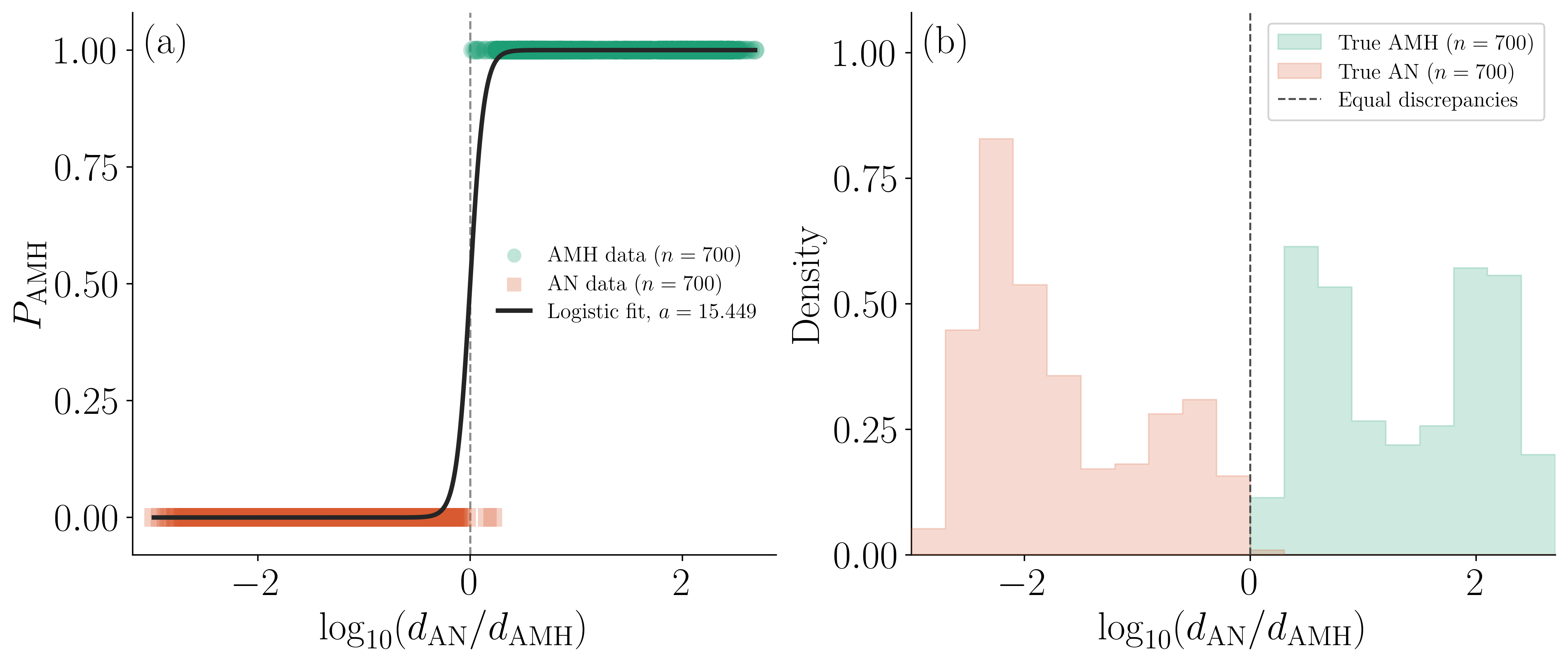}
    \caption{\textcolor{black}{\textbf{Model selection.} (a): plot of $P_{\mathrm{AMH}}=1-P_{\mathrm{AN}}$. See Eq.\eqref{eq:P_AMN_AN}. 
    The probability is plotted using the known model labels for 700 held-out AMH and 700 held-out AN simulations are plotted against
$\ell=\log_{10}(d_{\mathrm{AN}}/d_{\mathrm{AMH}})$; the solid curve is the fitted
$P_{\mathrm{AMH}}=[1+\exp(-a\ell)]^{-1}$ with $a=15.448875$.
See Eq.\eqref{eq:P_AMN_AN} of the text.
The markers are individual class labels, not probability estimates.
(b): density distributions of $\log_{10}(d_{\mathrm{AN}}/d_{\mathrm{AMH}})$
for the two true models. The dashed line marks equal discrepancies;
values to its right favor AMH and values to its left favor AN.}}
\label{fig:model_selection_cali}
\end{figure*}
%%================

Given an observed velocity field, we compute its energy spectrum and pass
it through both trained networks, obtaining candidate parameters
$\theta_{\mathrm{AMH}}$ and $\theta_{\mathrm{AN}}$. For each candidate, we
run new simulations using the predicted parameters, with $R$ independent
realizations to account for the randomness in the initial conditions.
Because the energy spectrum fluctuates in time even at fixed parameters, we
compare models using the time-averaged spectrum, $\log_{10}\langle
E(k,t)\rangle_t$, rather than comparing individual snapshots; averaging
over time removes fluctuations that are not specific to any one model and
retains the steady-state spectral shape that is characteristic of a given
parameter set. For each candidate model $c$, we define a discrepancy score $d_c$:
\begin{equation}
d_c = \frac{1}{R N_k}\sum_{r,k}
\Big[\log_{10}\langle E \rangle_t - \log_{10}
\langle \hat{E}^{(r)}\rangle_t 
\Big]^2.
\end{equation}
In the above, $d_c$ is the mean squared error, in log space, between the observed $\langle {E}\rangle_t $
and simulated
spectra $\langle \hat{E}^{(r)}\rangle_t$, averaged over the $R$ realizations and over wavenumber bins\cite{toni2008approximate}. 

{\color{black}
For each candidate model $c$, we compare its spectral discrepancy $d_c$ with a limit $T_c$ calibrated from held-out simulations generated by that model. The limit $T_c$ is the upper-tail cutoff corresponding to a significance
level of $0.01$: only about $1\%$ of held-out simulations from model $c$
have discrepancies above this value. We define the normalized discrepancy $G_c$ as: 
$$G_c=\frac{d_c}{T_c};
$$
a candidate passes this absolute-fit check when $G_c\leq 1$.  If neither candidate passes, we return NOTT (neither of the two). If only one passes, we select it; if both pass, we select the one with the smaller discrepancy. NOTT is therefore an outcome of the absolute-fit check, not a third model learned by the networks.

We separately estimate the probability of AMH \emph{conditional on choosing between AMH and AN}. We refer to it as $P({\mathrm{AMH}|\ell})=P_{\mathrm{AMH}}$. Explicitly, it is given as:
\begin{equation}
P_{\mathrm{AMH}}=\frac{1}{1+\exp(-a\ell)},
\qquad
P_{\mathrm{AN}}=1-P_{\mathrm{AMH}}.
\label{eq:P_AMN_AN}
\end{equation}
Here, we have defined:
\begin{align}
\ell=\log_{10}\!\left(\frac{d_{\mathrm{AN}}}
                    {d_{\mathrm{AMH}}}\right).    
\end{align}
The slope fitted using 700 held-out simulations of each model is $a=15.448875$. {\color{black}
Equation~\eqref{eq:P_AMN_AN} is an empirical way to convert the relative
discrepancy $\ell$ into a probability; it does not follow directly from the
continuum equations. We use a logistic curve because it changes smoothly
from favoring AN to favoring AMH and gives equal probabilities when the two
discrepancies are equal ($\ell=0$). We set the parameter $a$ by fitting the curve to
the known model labels of 1400 simulations. To assess the fitted relation, we used leave-one-out validation: each case
was omitted in turn, the curve was fitted to the remaining cases, and the
predicted probability for the omitted case was compared with its known label \cite{stone1974val}.  We quantified the probability prediction error using
the Brier score \cite{brier1950}. The fitting and scoring were performed with
\texttt{scikit-learn}~\cite{scikit-learn}.
}This probability measures relative support for AMH versus AN; it does not indicate whether either model passes the absolute-fit check. The individual model labels, fitted probability, and discrepancy-ratio
distributions are shown in Fig.~\ref{fig:model_selection_cali}.

{\color{black}
We calibrated the logistic slope and the absolute-fit limits using 1,400
simulations not used to train the networks: 700 AMH simulations and 700 AN
simulations. We then evaluated the fixed procedure on a separate set of 100 newly generated AMH observations and 100 newly generated AN observations. These 200 new test observations were not used to train the networks or to calibrate the logistic slope, absolute-fit limits, or parameter intervals. The results below are therefore separate from the 1,400 calibration cases.} All 100 AMH observations were assigned to AMH. Of the AN observations, 97 were assigned to AN, two to AMH, and one to NOTT. Fig.~\ref{fig:standard_absolute_fit} 
in appendix \ref{app:absoluteFit} shows both the relative discrepancy and the absolute-fit check for these observations. In particular, the NOTT outcome demonstrates that the procedure does not have to choose a model when both candidate spectra fit poorly.
}

%%

%%%%===========================
%%%%===========================
%%%%===========================
\section{Summary and discussion}
\label{sec:summary}
%%%%===========================
%%%%===========================
%%%%===========================

We introduce a machine-learning 
framework designed to solve the inverse problem of model selection and parameter estimation in continuum models of 
active matter. While forward modeling traditionally uses material parameters to simulate flow, mapping observable kinematics back to their generative, non-equilibrium forces represents a critical challenge. By treating high-resolution hydrodynamic velocity fields—such as those routinely captured via Particle Image Velocimetry (PIV)—as a diagnostic readout, our framework non-invasively reconstructs internal driving forces and infers underlying activity parameters. Thus, our framework \textcolor{black}{provides} new pathways for characterizing active matter systems and potential designing 
ideas for autonomous, smart soft materials.
We demonstrate the versatility and robustness of this approach across distinct classes of continuum descriptions, specifically Active Model H and Active Nematics. 
 
Our methodology consists of three stages: extracting the kinetic energy spectrum from velocity frames, training a convolutional neural network to predict activity parameters from kinetic energy spectrum, and using resimulation to select between competing models. A conceptual overview of this pipeline is given in Fig.~\ref{fig:network_architecture} and Algorithm~\ref{alg:modelS}. In this paper, we have considered model inference and selection of AMH and active nematics for two-dimensional systems where the model reduces to a scalar field \cite{alert2020universal}. Applying the method presented here to three-dimensional systems and models with more complexity suggests an exciting direction for future work. \textcolor{black}{The approach can be extended to experimental and biological flow data, where measurement noise, finite spatial and temporal resolution, and deviations from the assumed continuum descriptions will need to be considered.}

Machine learning methods have been used 
to predict the activity from fluid flow in
 particle-based models \cite{mohapatra2025inferring,bayati2025inferring}.
In this paper, we focus on continuum models and have chosen two representative models which have activity in the mechanical sector for our inference framework. Thus, the fluid flow retains a high-fidelity signature. Extending our method to cases when the activity is in the diffusive sector \cite{tjhung2018cluster, cates2019active}
suggest a direction for future work. In addition, we have considered a small range for the parameter inference to avoid computational cost. We believe that our model is extendable to larger systems sizes and data in three-dimensions. In Ref.\cite{colen2021machine} director field data was used for inference which was then used for forecasting future. In our method, we simply use the fluid velocity to obtain the parameters of the model. To find the corresponding dynamics of the order parameter, we need some initial condition which can then be evolved using the parameters. It should be noted that deep learning methodology of this paper is only expected to work for Stokesian fluid-flow data. For more complex scenarios - for example, cases where inertia and/or memory effects in the fluid are important – 
different architectures and strategies of inference and model selection are required.

\textcolor{black}{ The scope of the present study is restricted to 
two wet active-matter descriptions: Active Model H and the minimal active nematic model. We do not consider polar active fluids or dry active matter, both of which represent important classes of nonequilibrium systems. These theories may differ in their relevant order parameters, conservation laws, and hydrodynamic couplings. Extending the framework to these classes would require suitable forward models and training ensembles, together with validation procedures that distinguish relative model preference from absolute model adequacy. These results are therefore interpreted as a comparison between two distinct continuum descriptions, rather than as a broad classification scheme for active matter systems. In addition, the framework does not isolate the role of scalar versus tensorial order parameters. A more systematic study would require models in which density dynamics and orientational order can be varied independently, which is an interesting direction of future work.}

%%%+=======
%%%+=======
%%%+=======
\section*{Declaration of interests} The authors report no conflict of interest.
%%%+=======
%%%+=======
\section*{Data availability} All code used in this work is freely available on GitHub:
\href{https://github.com/soft-matter-physics/active-field-theories}{https://github.com/soft-matter-physics/active-field-theories}.
%%%+=======
%%%+=======
%%%+=======
\appendix
%%%+==================
%%%+==================

%%
%%

%%
%%
\section{Details of simulation method}\label{sec:appMethod}

In this section, we explain the numerical methods used to simulate continuum models of active matter with fluid flow.
At low Reynolds number, as applicable to active colloidal matter, 
the fluid flow satisfies the Stokes equation
\begin{gather}
-\boldsymbol{\nabla}p+\eta\nabla^{2}\mathbf{v}  =-\boldsymbol{f},\label{eq:stokes1}\\
\boldsymbol{\nabla}\cdot\mathbf{v} =0.
\label{eq:incomp}
\end{gather}
Here, $\boldsymbol{f}$
is the force density in the fluid which is due to activity of the medium. In the above, we ignore fluid inertia and thermal fluctuations.
The force density in the fluid is distinct for active model H of section 
 \ref{sec:amh}
and active nematics of section  \ref{sec:an}.
The fluid incompressibility is defined using Eq.\eqref{eq:incomp}.
In the following, we explain a numerical scheme which ensures incompressibility of the fluid exactly. 

The forward problem: to determine the flow
given the force density. 
We use Fourier transforms to obtain a solution of the
above equations. We define the Fourier transform of a function $\varphi(\boldsymbol{r})$ in $d-$dimensions
as\begin{subequations}\label{eq:FT} 
\begin{gather}
\hat{\varphi}(\boldsymbol{k})=\mathbb{\mathbb{F}}\left[\varphi(\boldsymbol{r})\right]=\int\varphi(\boldsymbol{r})\,e^{-i\boldsymbol{k}\cdot\boldsymbol{r}}\,d\boldsymbol{r},\\
\varphi(\boldsymbol{r})=\mathbb{\mathbb{F}}^{-1}\left[\hat{\varphi}(\boldsymbol{k})\right]=\frac{1}{(2\pi)^{d}}\int\hat{\varphi}(\boldsymbol{k})\,e^{i\boldsymbol{k}\cdot\boldsymbol{r}}\,d\boldsymbol{k}.
\end{gather}
\end{subequations}We now Fourier transform (\ref{eq:stokes1}\ref{eq:incomp}) to
obtain:
\begin{gather}
-i\boldsymbol{k}\hat{p}-\eta k^{2}\boldsymbol{\hat{v}} 
=-\boldsymbol{\hat{f}},\label{eq:stokes-k}
\\
i\boldsymbol{k}\cdot\boldsymbol{\hat{v}} =0.%\label{eq:incompressibility}
\end{gather}
The above equations can then be used to obtain
the Fourier transform of the pressure field (note that all isotropic terms have been absorbed in the pressure, which is then used to ensure the incompressibility condition):
%\begin{equation}\%end{equation}
$$
\hat{p}=-i\boldsymbol{k}\cdot\boldsymbol{\hat{f}}/k^{2}.
$$ 
The above expression of the pressure is then used in (\ref{eq:stokes-k})
to obtain the solution of the fluid flow, with built-in incompressibility,
given as 
\begin{align}
\boldsymbol{\hat{v}} & =\mathbf{\hat{G}}(\boldsymbol{k})\cdot\boldsymbol{\hat{f}},\quad\quad\mathbf{\hat{G}}(\boldsymbol{k})=\frac{1}{\eta}\left(\frac{\boldsymbol{I}}{k^{2}}-\frac{\boldsymbol{k}\boldsymbol{k}}{k^{4}}\right).\label{eq:stokes-k-1}
\end{align}
Here $\mathbf{\hat{G}}(\boldsymbol{k})$ is the Fourier transform
of the Oseen tensor $\mathbf{G}(\boldsymbol{r})$ \cite{pozrikidis1992}.
Using the above solution, the simulations are performed using the 
pseudo-spectral method
\cite{boyd2001chebyshev,kutz2026data}.
The solution for the fluid flow and order parameter is numerically
implemented using standard fast Fourier transforms (FFTs). Thus, 
periodic boundary conditions are automatically
ensured in the system and incompressibility condition is exactly satisfied. 
Throughout the paper, the spatial domain is discretized
using a uniform grid of $N_g = 128$ points along each direction, and $\eta = 1$. The values of $\tilde\kappa$ is taken from (-1, 0), while parameter $b$ varies from (0.25, 1.25) and parameter $K$ varies from $(0.5, 2)$. The activity parameter 
$\zeta$ for active nematics is taken from (-1.2, 0.1).
AI tools assisted with code development and debugging. The authors reviewed, tested, and validated all AI-suggested code modifications.

%%%+==================
\begin{figure}[ ]
    \includegraphics[width=0.88\columnwidth]{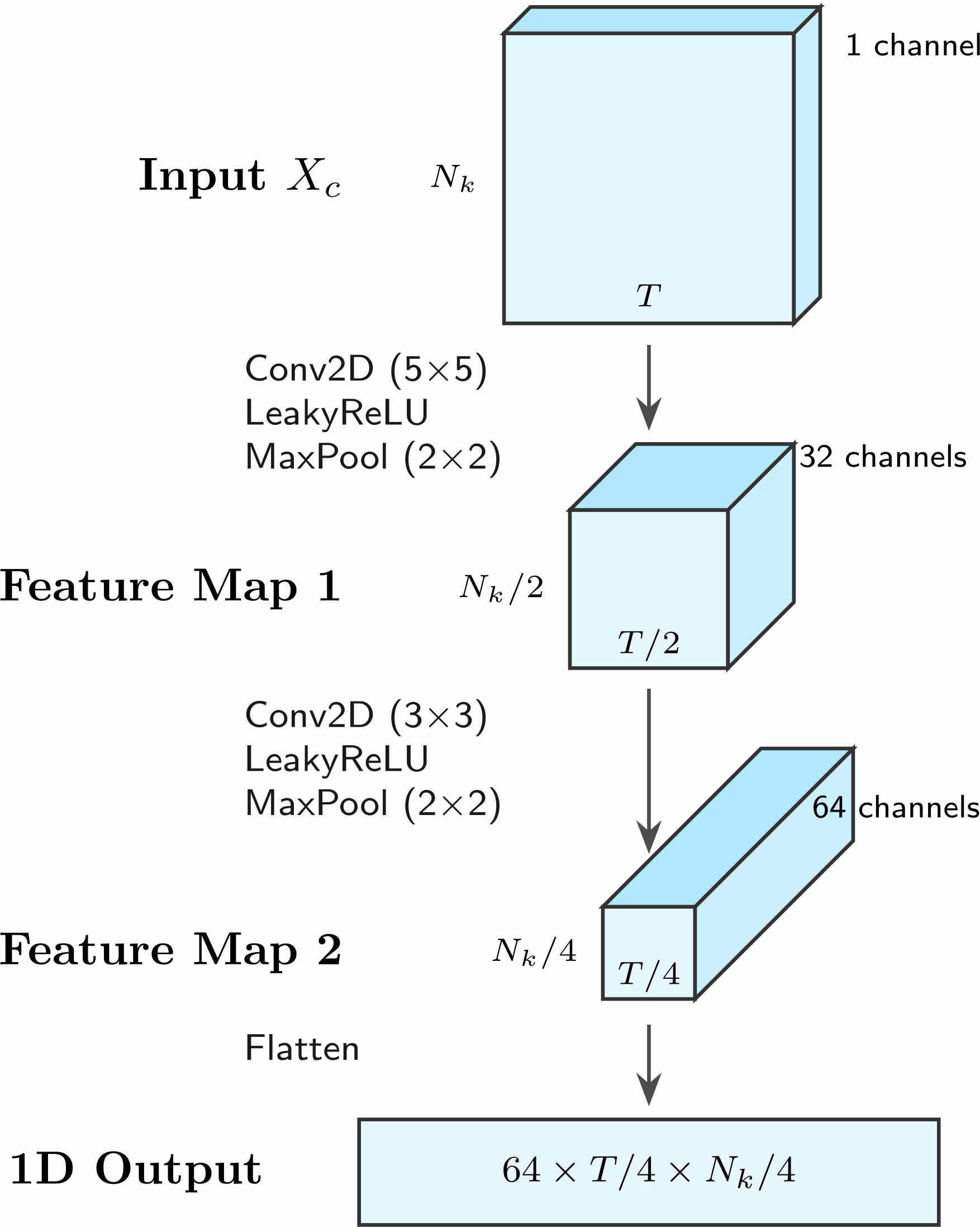}     
    \caption{Processioning of 
    the energy spectrum by the 2D Convolutional Neural Network.}
    \label{fig:cnn_architecture}
\end{figure}
%%%+==================

%================
\begin{figure*}[ ]
    \centering
    \includegraphics[width=0.99\linewidth]{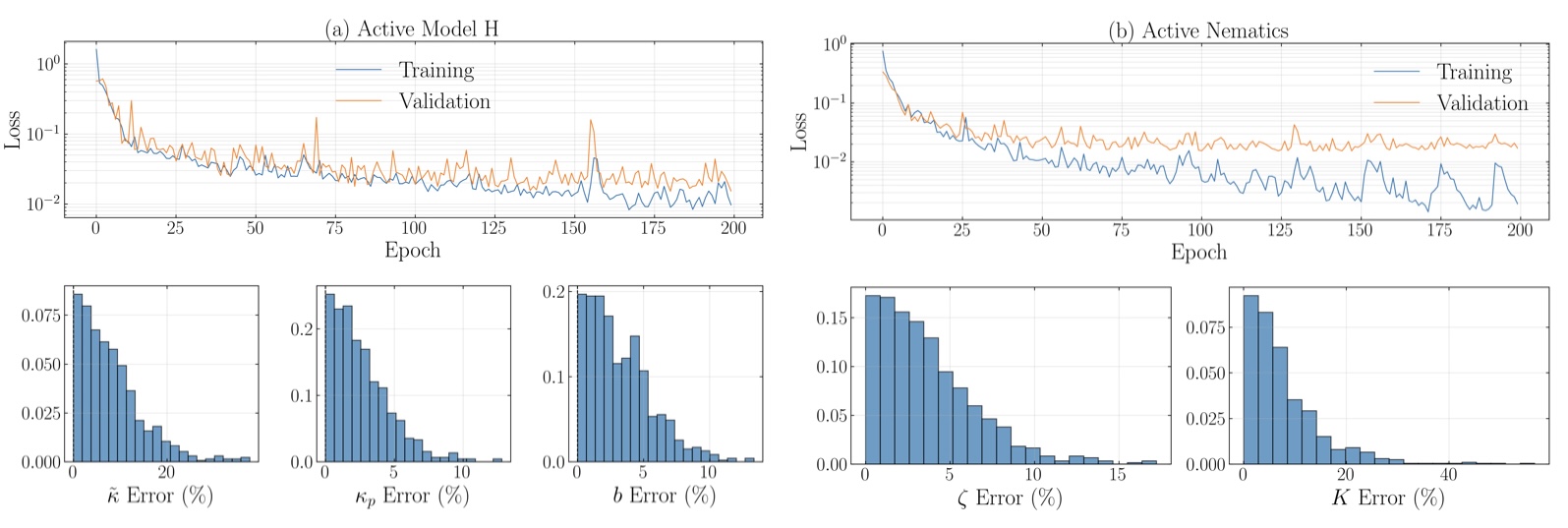}     
    \caption{(a) Convergence plot of Active Model H: Training and Validation loss versus epoch, $\tilde \kappa$ prediction percentage error, $\kappa_p$ prediction percentage error, and $b$ prediction percentage error. (b) Convergence plot of Active nematics model:  Training and Validation loss versus epoch, $\zeta$ prediction percentage error, and $K$ prediction percentage error.}
    \label{convergence_amh}
\end{figure*}
%%============
%%%+==================
\section{Details of inference framework}
%%%+==================
\subsection{The 2D Convolutional Neural Network}
The figure \ref{fig:cnn_architecture} shows how the 2D Convolutional Neural Network processes the energy spectrum. The network treats the 2D spectrum like an image, passing it through two sequential processing blocks. Each block uses a convolutional layer, followed by a pooling layer that reduces the spatial resolution. The feature maps become progressively smaller in the temporal and spectral dimensions, the number of channels increases, allowing the network to encode more abstract and discriminative patterns. The final feature map is reshaped into a one-dimensional vector, which serves as the input to subsequent stages of the model.

% \subsection{Algorithm}
% The algorithm used to select models is given in Fig.\ref{alg:modelS}
\subsection{Convergence plot}
The convergence of the ML framework is shown in Fig.\ref{convergence_amh}. 
It can be seen that training loss and validation loss decrease over epochs. Thus, it indicates that the model is learning well without
overfitting to the data.
 
 %
% \subsection{Energy spectrum}
% In Fig~\ref{Ek_amh}, plots of the kinetic energy spectrum $E(k)$ versus $k$ are shown for two different time instances $t=40000$ (left panel) and $t=80000$ (right panel) for Active Model H. Keeping the other parameters fixed, the variation of the energy spectrum with respect to the parameter $\tilde \kappa$ is shown in Fig~\ref{Ek_amh}(a)-(b). Similarly, the energy spectrum for different values of the parameters $b$ and $\kappa_p$ are shown in Fig~\ref{Ek_amh}(c)-(d) and (e)-(f), respectively.\\

% Next, we have shown the energy spectrum for the Active Nematics in Fig~\ref{Ek_an} 
% for two different time instances $t=40000$ (left panel) and $t=80000$ (right panel).
% For different values of the parameters $\zeta$ and $K$, $E(k)$ is plotted with $k$ and shown in Fig~\ref{Ek_an}(a)-(b) and (c)-(d) respectively.

%%%%%%%%%%%%%%%%%%%%%%%%%%%%%%%%%%%%%%%%%%%%%%%%%%%%%%%
% \begin{figure}%[h!]
%     \centering
%     \includegraphics[width=0.98\columnwidth]{Ek_an.jpeg}     
%     \caption{Active Nematics model: (a), (b) Energy spectrum plot at two different time instances with different values of $\zeta$. (c), (d) Energy spectrum at two different time instances with different values of $K$. }
%     \label{Ek_an}
% \end{figure}
%%%%%%%%%%%%%%%%%%%%%%%%%%%%%%%%%%%%%%%%%%%%%
\section{Absolute-fit check for standard AMH and AN data}
\label{app:absoluteFit}
\begin{figure*}[]
    \centering
    \includegraphics[width=0.75\textwidth]{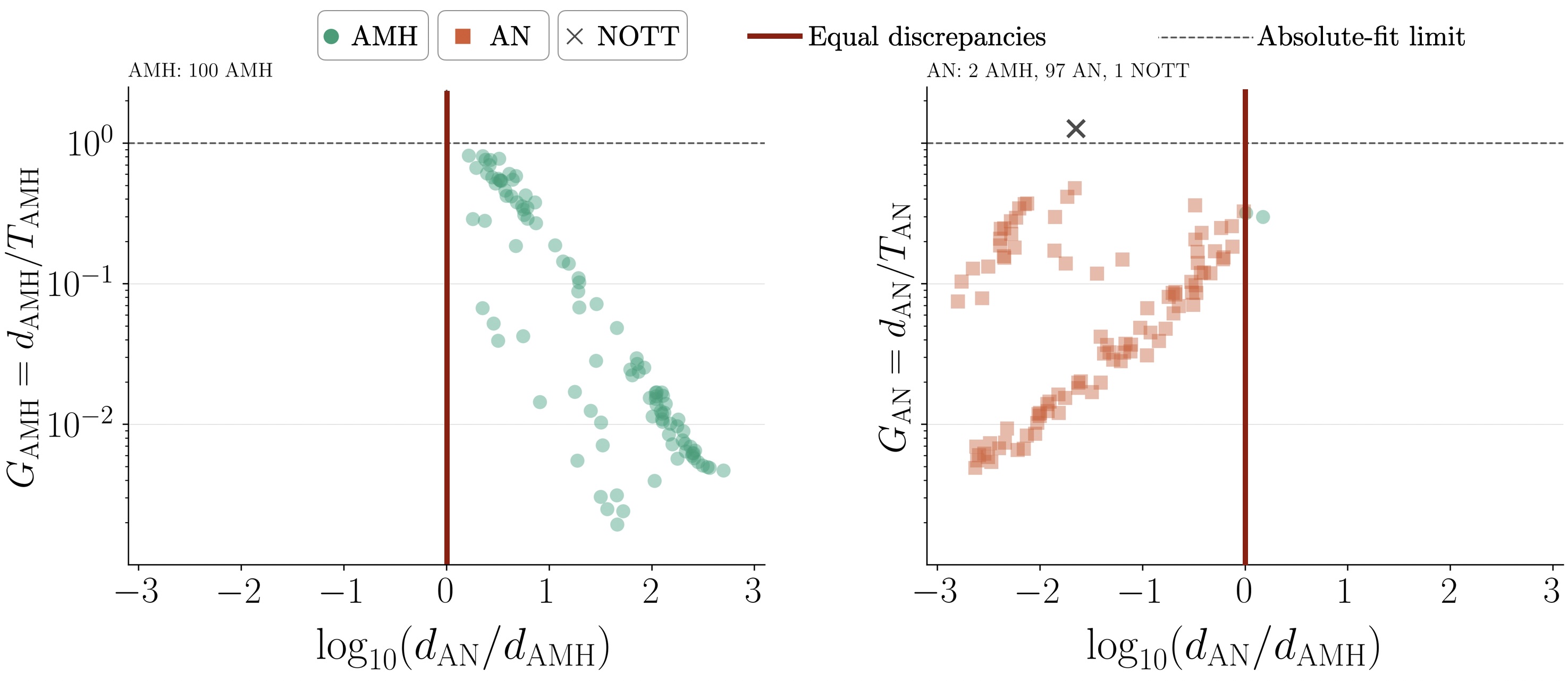}
    % {new_figs/standard_amh_an_relative_and_absolute_fit.png}
    \caption{\textcolor{black}{\textbf{Model preference and absolute spectral fit for standard AMH and AN observations.}
    Each panel contains 100 new observations generated by the model named in its title. Both panels use
    $\log_{10}(d_{\mathrm{AN}}/d_{\mathrm{AMH}})$ horizontally; positive values favor AMH and negative values favor AN.
    The vertical axis is $G_{\mathrm{AMH}}=d_{\mathrm{AMH}}/T_{\mathrm{AMH}}$ for true AMH observations (left) and
    $G_{\mathrm{AN}}=d_{\mathrm{AN}}/T_{\mathrm{AN}}$ for true AN observations (right).
    The thick and dashed vertical line (red) marks equal discrepancies, and the dashed horizontal line marks the absolute-fit limit $G_c=1$.
    Marker color and shape indicate the final decision: AMH, AN, or NOTT (neither of the two).
    The outcomes are 100 AMH decisions for true AMH data and, for true AN data, 97 AN decisions, two AMH decisions, and one NOTT decision. The test cases used are different from what was used for calibration.}}
    \label{fig:standard_absolute_fit}
\end{figure*}

{\color{black}
We first examine the absolute-fit check on new observations generated by the same AMH and AN equations used to develop the method. Figure~\ref{fig:standard_absolute_fit} plots the relative preference between the models against the normalized discrepancy of the observation's true model. This makes it possible to see separately whether the correct model reproduces the observed spectrum and which candidate is ultimately selected. An absolute-fit limit can also reject an occasional observation from a model that is present in the candidate set: one of the 100 standard AN observations receives the NOTT outcome.
}

\section{AMH and role of $\lambda$}
\label{app:lambda}

\begin{figure*}[]
    \centering
    \includegraphics[width=0.75\linewidth]{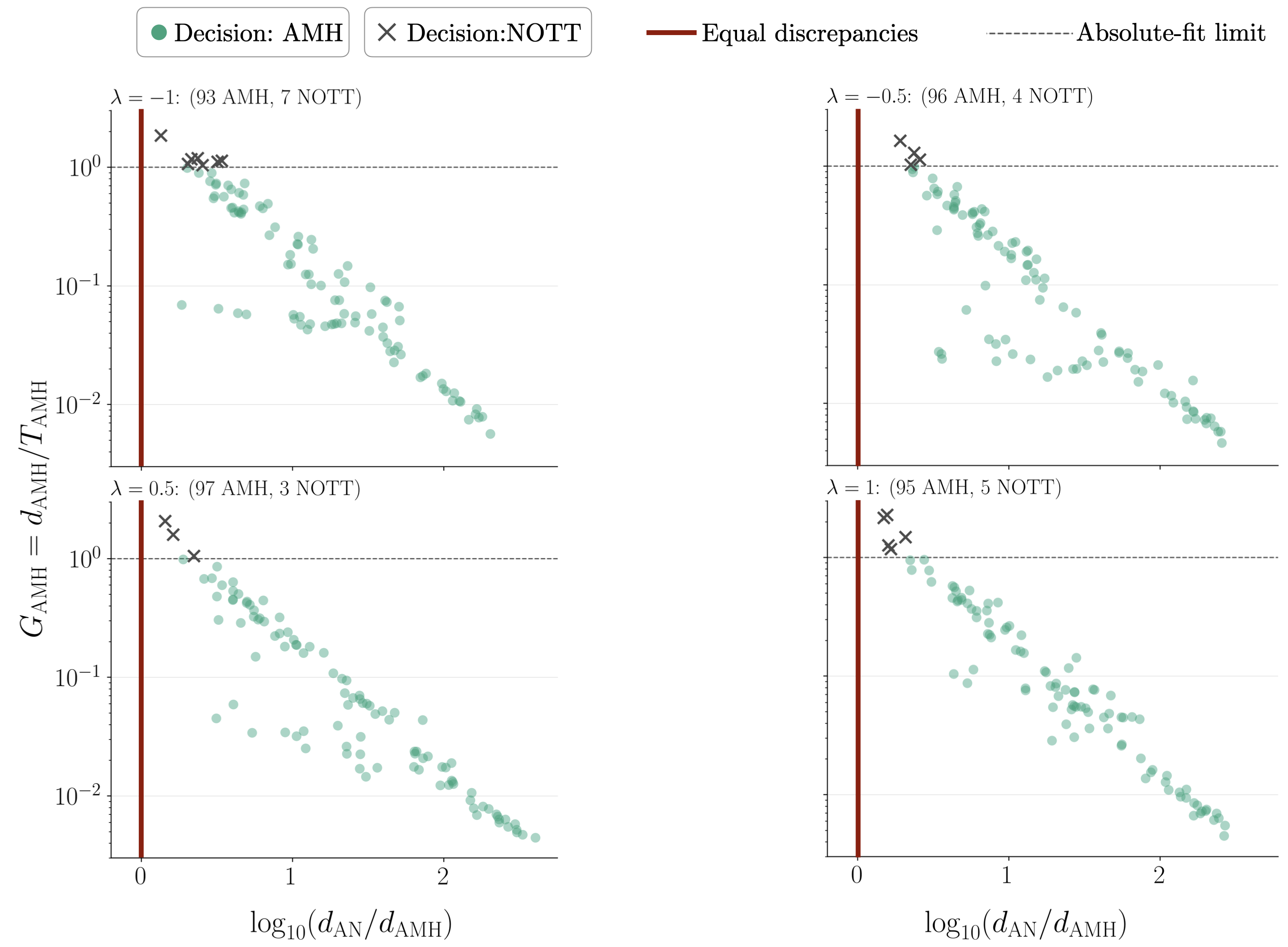}
    \caption{\textcolor{black}{\textbf{Model selection for modified AMH dynamics.}
    Each panel shows 100 observations generated at the indicated value of $\lambda$.
    The horizontal coordinate is the relative discrepancy
    $\log_{10}(d_{\mathrm{AN}}/d_{\mathrm{AMH}})$; values above zero favor AMH.
    The vertical coordinate is the normalized absolute AMH discrepancy
    $G_{\mathrm{AMH}}=d_{\mathrm{AMH}}/T_{\mathrm{AMH}}$, where
    $T_{\mathrm{AMH}}$ is the calibrated fit limit. The red 
    solid vertical line
    marks equal discrepancies, and the dashed horizontal line marks
    $G_{\mathrm{AMH}}=1$. Green circles are accepted as AMH; crosses are
    labelled NOTT (neither of the two candidates passes the absolute-fit
    check). All observations favor AMH over AN relatively, while
    7, 4, 3, and 5 observations, respectively, are rejected by the
    absolute-fit check.}}
    \label{fig:lambda_robustness}
\end{figure*}

{\color{black}To test sensitivity to dynamics absent from the training data, we generated new AMH observations with an additional $\lambda|\nabla\phi|^2$ term in the chemical potential. This term is a non-equilibrium addition to the chemical potential as it is not derived from any free energy \cite{wittkowski2014scalar}. 
The trained networks were left unchanged. We evaluated 100 observations at each nonzero value of $\lambda$ shown in Fig.~\ref{fig:lambda_robustness}.
The machine learning model is still able to effectively learn parameters and models. 
% In addition, we can add an non-equilibrium term to the current \cite{tjhung2018cluster}. 
}

\section{Empirical parameter-uncertainty intervals}
\label{app:param_uncertainty}

{\color{black}
For each parameter, we construct an empirical $90\%$ interval from
parameter-recovery errors on the 1400 held-out calibration simulations of AMH and AN. For a
predicted parameter magnitude $|\hat{\theta}_j|$, the interval is
\begin{equation}
\mathcal I_j=
\left[
\frac{|\hat{\theta}_j|}{f_j},
|\hat{\theta}_j|f_j
\right],
\end{equation}
where $f_j$ is determined from the $90$th percentile of the absolute errors
in $\log_{10}|\theta_j|$. For parameters whose signs are fixed by convention,
the interval is formed for the magnitude and the prescribed sign is restored.

Table~\ref{tab:param_uncertainty} reports the multiplicative factors and the
fraction of true parameter values covered in separate standard test sets of
100 AMH and 100 AN observations which were not used for calibration. The observed coverage is close to $90\%$ for these test sets. These results apply to the simulation conditions tested here.
}

\begin{table}[t]
\caption{Multiplicative factors for the empirical $90\%$ parameter intervals
and observed coverage on separate standard test data.}
\label{tab:param_uncertainty}
\begin{ruledtabular}
\begin{tabular}{llcc}
Model & Parameter & Factor $f_j$ & Coverage\\
\hline
AMH & $|\tilde{\kappa}|$ & 1.186 & 88\%\\
AMH & $\kappa_p$         & 1.057 & 91\%\\
AMH & $b$                & 1.068 & 91\%\\
AN  & $\kappa$           & 1.180 & 88\%\\
AN  & $|\zeta|$          & 1.078 & 87\%\\
\end{tabular}
\end{ruledtabular}
\end{table}

%A list of different parameter values used for the analysis (figure) is given in Table \ref{si_table_params}.
%%=======================
%%=======================
% \bibliography{references}

\begin{thebibliography}{50}%
\makeatletter
\providecommand \@ifxundefined [1]{%
 \@ifx{#1\undefined}
}%
\providecommand \@ifnum [1]{%
 \ifnum #1\expandafter \@firstoftwo
 \else \expandafter \@secondoftwo
 \fi
}%
\providecommand \@ifx [1]{%
 \ifx #1\expandafter \@firstoftwo
 \else \expandafter \@secondoftwo
 \fi
}%
\providecommand \natexlab [1]{#1}%
\providecommand \enquote  [1]{``#1''}%
\providecommand \bibnamefont  [1]{#1}%
\providecommand \bibfnamefont [1]{#1}%
\providecommand \citenamefont [1]{#1}%
\providecommand \href@noop [0]{\@secondoftwo}%
\providecommand \href [0]{\begingroup \@sanitize@url \@href}%
\providecommand \@href[1]{\@@startlink{#1}\@@href}%
\providecommand \@@href[1]{\endgroup#1\@@endlink}%
\providecommand \@sanitize@url [0]{\catcode `\\12\catcode `\$12\catcode `\&12\catcode `\#12\catcode `\^12\catcode `\_12\catcode `\%12\relax}%
\providecommand \@@startlink[1]{}%
\providecommand \@@endlink[0]{}%
\providecommand \url  [0]{\begingroup\@sanitize@url \@url }%
\providecommand \@url [1]{\endgroup\@href {#1}{\urlprefix }}%
\providecommand \urlprefix  [0]{URL }%
\providecommand \Eprint [0]{\href }%
\providecommand \doibase [0]{https://doi.org/}%
\providecommand \selectlanguage [0]{\@gobble}%
\providecommand \bibinfo  [0]{\@secondoftwo}%
\providecommand \bibfield  [0]{\@secondoftwo}%
\providecommand \translation [1]{[#1]}%
\providecommand \BibitemOpen [0]{}%
\providecommand \bibitemStop [0]{}%
\providecommand \bibitemNoStop [0]{.\EOS\space}%
\providecommand \EOS [0]{\spacefactor3000\relax}%
\providecommand \BibitemShut  [1]{\csname bibitem#1\endcsname}%
\let\auto@bib@innerbib\@empty
%</preamble>
\bibitem [{\citenamefont {Cates}\ and\ \citenamefont {MacKintosh}(2011)}]{cates2011}%
  \BibitemOpen
  \bibfield  {author} {\bibinfo {author} {\bibfnamefont {M.~E.}\ \bibnamefont {Cates}}\ and\ \bibinfo {author} {\bibfnamefont {F.~C.}\ \bibnamefont {MacKintosh}},\ }\bibfield  {title} {\bibinfo {title} {{Active soft matter}},\ }\href {https://doi.org/10.1039/c1sm90014e} {\bibfield  {journal} {\bibinfo  {journal} {Soft Matter}\ }\textbf {\bibinfo {volume} {7}},\ \bibinfo {pages} {3050} (\bibinfo {year} {2011})}\BibitemShut {NoStop}%
\bibitem [{\citenamefont {Ramaswamy}(2010)}]{ramaswamy2010}%
  \BibitemOpen
  \bibfield  {author} {\bibinfo {author} {\bibfnamefont {S.}~\bibnamefont {Ramaswamy}},\ }\bibfield  {title} {\bibinfo {title} {{The Mechanics and Statistics of Active Matter}},\ }\href {https://doi.org/10.1146/annurev-conmatphys-070909-104101} {\bibfield  {journal} {\bibinfo  {journal} {Annu. Rev. Condens. Mat. Phys.}\ }\textbf {\bibinfo {volume} {1}},\ \bibinfo {pages} {323} (\bibinfo {year} {2010})}\BibitemShut {NoStop}%
\bibitem [{\citenamefont {Marchetti}\ \emph {et~al.}(2013)\citenamefont {Marchetti}, \citenamefont {Joanny}, \citenamefont {Ramaswamy}, \citenamefont {Liverpool}, \citenamefont {Prost}, \citenamefont {Rao},\ and\ \citenamefont {Simha}}]{marchetti2013}%
  \BibitemOpen
  \bibfield  {author} {\bibinfo {author} {\bibfnamefont {M.~C.}\ \bibnamefont {Marchetti}}, \bibinfo {author} {\bibfnamefont {J.~F.}\ \bibnamefont {Joanny}}, \bibinfo {author} {\bibfnamefont {S.}~\bibnamefont {Ramaswamy}}, \bibinfo {author} {\bibfnamefont {T.~B.}\ \bibnamefont {Liverpool}}, \bibinfo {author} {\bibfnamefont {J.}~\bibnamefont {Prost}}, \bibinfo {author} {\bibfnamefont {M.}~\bibnamefont {Rao}},\ and\ \bibinfo {author} {\bibfnamefont {R.~A.}\ \bibnamefont {Simha}},\ }\bibfield  {title} {\bibinfo {title} {Hydrodynamics of soft active matter},\ }\href {https://doi.org/10.1103/RevModPhys.85.1143} {\bibfield  {journal} {\bibinfo  {journal} {Rev. Mod. Phys.}\ }\textbf {\bibinfo {volume} {85}},\ \bibinfo {pages} {1143} (\bibinfo {year} {2013})}\BibitemShut {NoStop}%
\bibitem [{\citenamefont {te~Vrugt}\ \emph {et~al.}(2026)\citenamefont {te~Vrugt}, \citenamefont {Liebchen},\ and\ \citenamefont {Cates}}]{te2026colloquium}%
  \BibitemOpen
  \bibfield  {author} {\bibinfo {author} {\bibfnamefont {M.}~\bibnamefont {te~Vrugt}}, \bibinfo {author} {\bibfnamefont {B.}~\bibnamefont {Liebchen}},\ and\ \bibinfo {author} {\bibfnamefont {M.~E.}\ \bibnamefont {Cates}},\ }\bibfield  {title} {\bibinfo {title} {Colloquium: What do we mean by ‘active matter’?},\ }\href@noop {} {\bibfield  {journal} {\bibinfo  {journal} {Rev. Mod. Phys.}\ }\textbf {\bibinfo {volume} {98}},\ \bibinfo {pages} {031001} (\bibinfo {year} {2026})}\BibitemShut {NoStop}%
\bibitem [{\citenamefont {Cates}\ and\ \citenamefont {Nardini}(2025)}]{cates2025active}%
  \BibitemOpen
  \bibfield  {author} {\bibinfo {author} {\bibfnamefont {M.~E.}\ \bibnamefont {Cates}}\ and\ \bibinfo {author} {\bibfnamefont {C.}~\bibnamefont {Nardini}},\ }\bibfield  {title} {\bibinfo {title} {Active phase separation: new phenomenology from non-equilibrium physics},\ }\href@noop {} {\bibfield  {journal} {\bibinfo  {journal} {Rep. Prog. Phys.}\ }\textbf {\bibinfo {volume} {88}},\ \bibinfo {pages} {056601} (\bibinfo {year} {2025})}\BibitemShut {NoStop}%
\bibitem [{\citenamefont {Simha}\ and\ \citenamefont {Ramaswamy}(2002)}]{simha2002a}%
  \BibitemOpen
  \bibfield  {author} {\bibinfo {author} {\bibfnamefont {R.~A.}\ \bibnamefont {Simha}}\ and\ \bibinfo {author} {\bibfnamefont {S.}~\bibnamefont {Ramaswamy}},\ }\bibfield  {title} {\bibinfo {title} {{Hydrodynamic Fluctuations and Instabilities in Ordered Suspensions of Self-Propelled Particles}},\ }\href {https://doi.org/10.1103/PhysRevLett.89.058101} {\bibfield  {journal} {\bibinfo  {journal} {Phys. Rev. Lett.}\ }\textbf {\bibinfo {volume} {89}},\ \bibinfo {pages} {058101} (\bibinfo {year} {2002})}\BibitemShut {NoStop}%
\bibitem [{\citenamefont {Alert}\ \emph {et~al.}(2022)\citenamefont {Alert}, \citenamefont {Casademunt},\ and\ \citenamefont {Joanny}}]{alert2022active}%
  \BibitemOpen
  \bibfield  {author} {\bibinfo {author} {\bibfnamefont {R.}~\bibnamefont {Alert}}, \bibinfo {author} {\bibfnamefont {J.}~\bibnamefont {Casademunt}},\ and\ \bibinfo {author} {\bibfnamefont {J.-F.}\ \bibnamefont {Joanny}},\ }\bibfield  {title} {\bibinfo {title} {Active turbulence},\ }\href@noop {} {\bibfield  {journal} {\bibinfo  {journal} {Annu. Rev. Condens. Matter Phys.}\ }\textbf {\bibinfo {volume} {13}},\ \bibinfo {pages} {143} (\bibinfo {year} {2022})}\BibitemShut {NoStop}%
\bibitem [{\citenamefont {Colen}\ \emph {et~al.}(2021)\citenamefont {Colen}, \citenamefont {Han}, \citenamefont {Zhang}, \citenamefont {Redford}, \citenamefont {Lemma}, \citenamefont {Morgan}, \citenamefont {Ruijgrok}, \citenamefont {Adkins}, \citenamefont {Bryant}, \citenamefont {Dogic} \emph {et~al.}}]{colen2021machine}%
  \BibitemOpen
  \bibfield  {author} {\bibinfo {author} {\bibfnamefont {J.}~\bibnamefont {Colen}}, \bibinfo {author} {\bibfnamefont {M.}~\bibnamefont {Han}}, \bibinfo {author} {\bibfnamefont {R.}~\bibnamefont {Zhang}}, \bibinfo {author} {\bibfnamefont {S.~A.}\ \bibnamefont {Redford}}, \bibinfo {author} {\bibfnamefont {L.~M.}\ \bibnamefont {Lemma}}, \bibinfo {author} {\bibfnamefont {L.}~\bibnamefont {Morgan}}, \bibinfo {author} {\bibfnamefont {P.~V.}\ \bibnamefont {Ruijgrok}}, \bibinfo {author} {\bibfnamefont {R.}~\bibnamefont {Adkins}}, \bibinfo {author} {\bibfnamefont {Z.}~\bibnamefont {Bryant}}, \bibinfo {author} {\bibfnamefont {Z.}~\bibnamefont {Dogic}}, \emph {et~al.},\ }\bibfield  {title} {\bibinfo {title} {Machine learning active-nematic hydrodynamics},\ }\href@noop {} {\bibfield  {journal} {\bibinfo  {journal} {Proc. Natl. Acad. Sci. U.S.A.}\ }\textbf {\bibinfo {volume} {118}},\ \bibinfo {pages} {e2016708118} (\bibinfo {year} {2021})}\BibitemShut {NoStop}%
\bibitem [{\citenamefont {Joshi}\ \emph {et~al.}(2022)\citenamefont {Joshi}, \citenamefont {Ray}, \citenamefont {Lemma}, \citenamefont {Varghese}, \citenamefont {Sharp}, \citenamefont {Dogic}, \citenamefont {Baskaran},\ and\ \citenamefont {Hagan}}]{joshi2022}%
  \BibitemOpen
  \bibfield  {author} {\bibinfo {author} {\bibfnamefont {C.}~\bibnamefont {Joshi}}, \bibinfo {author} {\bibfnamefont {S.}~\bibnamefont {Ray}}, \bibinfo {author} {\bibfnamefont {L.~M.}\ \bibnamefont {Lemma}}, \bibinfo {author} {\bibfnamefont {M.}~\bibnamefont {Varghese}}, \bibinfo {author} {\bibfnamefont {G.}~\bibnamefont {Sharp}}, \bibinfo {author} {\bibfnamefont {Z.}~\bibnamefont {Dogic}}, \bibinfo {author} {\bibfnamefont {A.}~\bibnamefont {Baskaran}},\ and\ \bibinfo {author} {\bibfnamefont {M.~F.}\ \bibnamefont {Hagan}},\ }\bibfield  {title} {\bibinfo {title} {Data-driven discovery of active nematic hydrodynamics},\ }\href {https://doi.org/10.1103/PhysRevLett.129.258001} {\bibfield  {journal} {\bibinfo  {journal} {Phys. Rev. Lett.}\ }\textbf {\bibinfo {volume} {129}},\ \bibinfo {pages} {258001} (\bibinfo {year} {2022})}\BibitemShut {NoStop}%
\bibitem [{\citenamefont {Zhou}\ \emph {et~al.}(2021)\citenamefont {Zhou}, \citenamefont {Joshi}, \citenamefont {Liu}, \citenamefont {Norton}, \citenamefont {Lemma}, \citenamefont {Dogic}, \citenamefont {Hagan}, \citenamefont {Fraden},\ and\ \citenamefont {Hong}}]{zhou2021}%
  \BibitemOpen
  \bibfield  {author} {\bibinfo {author} {\bibfnamefont {Z.}~\bibnamefont {Zhou}}, \bibinfo {author} {\bibfnamefont {C.}~\bibnamefont {Joshi}}, \bibinfo {author} {\bibfnamefont {R.}~\bibnamefont {Liu}}, \bibinfo {author} {\bibfnamefont {M.~M.}\ \bibnamefont {Norton}}, \bibinfo {author} {\bibfnamefont {L.}~\bibnamefont {Lemma}}, \bibinfo {author} {\bibfnamefont {Z.}~\bibnamefont {Dogic}}, \bibinfo {author} {\bibfnamefont {M.~F.}\ \bibnamefont {Hagan}}, \bibinfo {author} {\bibfnamefont {S.}~\bibnamefont {Fraden}},\ and\ \bibinfo {author} {\bibfnamefont {P.}~\bibnamefont {Hong}},\ }\bibfield  {title} {\bibinfo {title} {Machine learning forecasting of active nematics},\ }\href {https://doi.org/10.1039/d0sm01316a} {\bibfield  {journal} {\bibinfo  {journal} {Soft Matter}\ }\textbf {\bibinfo {volume} {17}},\ \bibinfo {pages} {738} (\bibinfo {year} {2021})}\BibitemShut {NoStop}%
\bibitem [{\citenamefont {Zaplotnik}\ \emph {et~al.}(2023)\citenamefont {Zaplotnik}, \citenamefont {Pi{\v{s}}ljar}, \citenamefont {{\v{S}}karabot},\ and\ \citenamefont {Ravnik}}]{zaplotnik2023neural}%
  \BibitemOpen
  \bibfield  {author} {\bibinfo {author} {\bibfnamefont {J.}~\bibnamefont {Zaplotnik}}, \bibinfo {author} {\bibfnamefont {J.}~\bibnamefont {Pi{\v{s}}ljar}}, \bibinfo {author} {\bibfnamefont {M.}~\bibnamefont {{\v{S}}karabot}},\ and\ \bibinfo {author} {\bibfnamefont {M.}~\bibnamefont {Ravnik}},\ }\bibfield  {title} {\bibinfo {title} {Neural networks determination of material elastic constants and structures in nematic complex fluids},\ }\href@noop {} {\bibfield  {journal} {\bibinfo  {journal} {Sci. Rep.}\ }\textbf {\bibinfo {volume} {13}},\ \bibinfo {pages} {6028} (\bibinfo {year} {2023})}\BibitemShut {NoStop}%
\bibitem [{\citenamefont {Piven}\ \emph {et~al.}(2024)\citenamefont {Piven}, \citenamefont {Darmoroz}, \citenamefont {Skorb},\ and\ \citenamefont {Orlova}}]{piven2024}%
  \BibitemOpen
  \bibfield  {author} {\bibinfo {author} {\bibfnamefont {A.}~\bibnamefont {Piven}}, \bibinfo {author} {\bibfnamefont {D.}~\bibnamefont {Darmoroz}}, \bibinfo {author} {\bibfnamefont {E.}~\bibnamefont {Skorb}},\ and\ \bibinfo {author} {\bibfnamefont {T.}~\bibnamefont {Orlova}},\ }\bibfield  {title} {\bibinfo {title} {Machine learning methods for liquid crystal research: phases, textures, defects and physical properties},\ }\href {https://doi.org/10.1039/d3sm01634j} {\bibfield  {journal} {\bibinfo  {journal} {Soft Matter}\ }\textbf {\bibinfo {volume} {20}},\ \bibinfo {pages} {1380} (\bibinfo {year} {2024})}\BibitemShut {NoStop}%
\bibitem [{\citenamefont {Frishman}\ and\ \citenamefont {Keren}(2021)}]{frishman2021learning}%
  \BibitemOpen
  \bibfield  {author} {\bibinfo {author} {\bibfnamefont {A.}~\bibnamefont {Frishman}}\ and\ \bibinfo {author} {\bibfnamefont {K.}~\bibnamefont {Keren}},\ }\bibfield  {title} {\bibinfo {title} {Learning active nematics one step at a time},\ }\href@noop {} {\bibfield  {journal} {\bibinfo  {journal} {Proc. Natl. Acad. Sci. U.S.A.}\ }\textbf {\bibinfo {volume} {118}},\ \bibinfo {pages} {e2102169118} (\bibinfo {year} {2021})}\BibitemShut {NoStop}%
\bibitem [{\citenamefont {Golden}\ \emph {et~al.}(2023)\citenamefont {Golden}, \citenamefont {Grigoriev}, \citenamefont {Nambisan},\ and\ \citenamefont {Fernandez-Nieves}}]{golden2023physically}%
  \BibitemOpen
  \bibfield  {author} {\bibinfo {author} {\bibfnamefont {M.}~\bibnamefont {Golden}}, \bibinfo {author} {\bibfnamefont {R.~O.}\ \bibnamefont {Grigoriev}}, \bibinfo {author} {\bibfnamefont {J.}~\bibnamefont {Nambisan}},\ and\ \bibinfo {author} {\bibfnamefont {A.}~\bibnamefont {Fernandez-Nieves}},\ }\bibfield  {title} {\bibinfo {title} {Physically informed data-driven modeling of active nematics},\ }\href@noop {} {\bibfield  {journal} {\bibinfo  {journal} {Sci. Adv.}\ }\textbf {\bibinfo {volume} {9}},\ \bibinfo {pages} {eabq6120} (\bibinfo {year} {2023})}\BibitemShut {NoStop}%
\bibitem [{\citenamefont {Tiribocchi}\ \emph {et~al.}(2015)\citenamefont {Tiribocchi}, \citenamefont {Wittkowski}, \citenamefont {Marenduzzo},\ and\ \citenamefont {Cates}}]{tiribocchi2015active}%
  \BibitemOpen
  \bibfield  {author} {\bibinfo {author} {\bibfnamefont {A.}~\bibnamefont {Tiribocchi}}, \bibinfo {author} {\bibfnamefont {R.}~\bibnamefont {Wittkowski}}, \bibinfo {author} {\bibfnamefont {D.}~\bibnamefont {Marenduzzo}},\ and\ \bibinfo {author} {\bibfnamefont {M.~E.}\ \bibnamefont {Cates}},\ }\bibfield  {title} {\bibinfo {title} {Active model {H}: scalar active matter in a momentum-conserving fluid},\ }\href {https://doi.org/10.1103/PhysRevLett.115.188302} {\bibfield  {journal} {\bibinfo  {journal} {Phys. Rev. Lett.}\ }\textbf {\bibinfo {volume} {115}},\ \bibinfo {pages} {188302} (\bibinfo {year} {2015})}\BibitemShut {NoStop}%
\bibitem [{\citenamefont {Singh}\ and\ \citenamefont {Cates}(2019)}]{singh2019amh}%
  \BibitemOpen
  \bibfield  {author} {\bibinfo {author} {\bibfnamefont {R.}~\bibnamefont {Singh}}\ and\ \bibinfo {author} {\bibfnamefont {M.~E.}\ \bibnamefont {Cates}},\ }\bibfield  {title} {\bibinfo {title} {Hydrodynamically interrupted droplet growth in scalar active matter},\ }\href {https://doi.org/10.1103/PhysRevLett.123.148005} {\bibfield  {journal} {\bibinfo  {journal} {Phys. Rev. Lett.}\ }\textbf {\bibinfo {volume} {123}},\ \bibinfo {pages} {148005} (\bibinfo {year} {2019})}\BibitemShut {NoStop}%
\bibitem [{\citenamefont {Cates}(2019)}]{cates2019active}%
  \BibitemOpen
  \bibfield  {author} {\bibinfo {author} {\bibfnamefont {M.~E.}\ \bibnamefont {Cates}},\ }\bibfield  {title} {\bibinfo {title} {Active field theories},\ }\href@noop {} {\bibfield  {journal} {\bibinfo  {journal} {arXiv preprint arXiv:1904.01330}\ } (\bibinfo {year} {2019})}\BibitemShut {NoStop}%
\bibitem [{\citenamefont {Shankar}\ \emph {et~al.}(2022)\citenamefont {Shankar}, \citenamefont {Souslov}, \citenamefont {Bowick}, \citenamefont {Marchetti},\ and\ \citenamefont {Vitelli}}]{shankar2022topological}%
  \BibitemOpen
  \bibfield  {author} {\bibinfo {author} {\bibfnamefont {S.}~\bibnamefont {Shankar}}, \bibinfo {author} {\bibfnamefont {A.}~\bibnamefont {Souslov}}, \bibinfo {author} {\bibfnamefont {M.~J.}\ \bibnamefont {Bowick}}, \bibinfo {author} {\bibfnamefont {M.~C.}\ \bibnamefont {Marchetti}},\ and\ \bibinfo {author} {\bibfnamefont {V.}~\bibnamefont {Vitelli}},\ }\bibfield  {title} {\bibinfo {title} {Topological active matter},\ }\href@noop {} {\bibfield  {journal} {\bibinfo  {journal} {Nat. Rev. Phys.}\ }\textbf {\bibinfo {volume} {4}},\ \bibinfo {pages} {380} (\bibinfo {year} {2022})}\BibitemShut {NoStop}%
\bibitem [{\citenamefont {Ramaswamy}(2017)}]{ramaswamy2017active}%
  \BibitemOpen
  \bibfield  {author} {\bibinfo {author} {\bibfnamefont {S.}~\bibnamefont {Ramaswamy}},\ }\bibfield  {title} {\bibinfo {title} {Active matter},\ }\href {http://iopscience.iop.org/article/10.1088/1742-5468/aa6bc5/meta} {\bibfield  {journal} {\bibinfo  {journal} {J. Stat. Mech.}\ }\textbf {\bibinfo {volume} {2017}},\ \bibinfo {pages} {054002} (\bibinfo {year} {2017})}\BibitemShut {NoStop}%
\bibitem [{\citenamefont {Hohenberg}\ and\ \citenamefont {Halperin}(1977)}]{hohenberg1977theory}%
  \BibitemOpen
  \bibfield  {author} {\bibinfo {author} {\bibfnamefont {P.~C.}\ \bibnamefont {Hohenberg}}\ and\ \bibinfo {author} {\bibfnamefont {B.~I.}\ \bibnamefont {Halperin}},\ }\bibfield  {title} {\bibinfo {title} {Theory of dynamic critical phenomena},\ }\href {https://doi.org/10.1103/RevModPhys.49.435} {\bibfield  {journal} {\bibinfo  {journal} {Rev. of Mod. Phys.}\ }\textbf {\bibinfo {volume} {49}},\ \bibinfo {pages} {435} (\bibinfo {year} {1977})}\BibitemShut {NoStop}%
\bibitem [{\citenamefont {Padhan}\ and\ \citenamefont {Pandit}(2025)}]{padhan2025cahn}%
  \BibitemOpen
  \bibfield  {author} {\bibinfo {author} {\bibfnamefont {N.~B.}\ \bibnamefont {Padhan}}\ and\ \bibinfo {author} {\bibfnamefont {R.}~\bibnamefont {Pandit}},\ }\bibfield  {title} {\bibinfo {title} {The cahn--hilliard--navier--stokes framework for multiphase fluid flows: laminar, turbulent and active},\ }\href@noop {} {\bibfield  {journal} {\bibinfo  {journal} {J. Fluid Mech.}\ }\textbf {\bibinfo {volume} {1010}},\ \bibinfo {pages} {P1} (\bibinfo {year} {2025})}\BibitemShut {NoStop}%
\bibitem [{\citenamefont {Radhakrishnan}\ \emph {et~al.}(2026)\citenamefont {Radhakrishnan}, \citenamefont {Serafin}, \citenamefont {Schmidt},\ and\ \citenamefont {Fodor}}]{radhakrishnan2026irreversibility}%
  \BibitemOpen
  \bibfield  {author} {\bibinfo {author} {\bibfnamefont {B.~N.}\ \bibnamefont {Radhakrishnan}}, \bibinfo {author} {\bibfnamefont {F.}~\bibnamefont {Serafin}}, \bibinfo {author} {\bibfnamefont {T.~L.}\ \bibnamefont {Schmidt}},\ and\ \bibinfo {author} {\bibfnamefont {{\'E}.}~\bibnamefont {Fodor}},\ }\bibfield  {title} {\bibinfo {title} {Irreversibility in scalar active turbulence: the role of topological defects},\ }\href@noop {} {\bibfield  {journal} {\bibinfo  {journal} {New J. Phys.}\ }\textbf {\bibinfo {volume} {28}},\ \bibinfo {pages} {034601} (\bibinfo {year} {2026})}\BibitemShut {NoStop}%
\bibitem [{\citenamefont {Landau}\ and\ \citenamefont {Lifshitz}(1959)}]{landau1959fluid}%
  \BibitemOpen
  \bibfield  {author} {\bibinfo {author} {\bibfnamefont {L.~D.}\ \bibnamefont {Landau}}\ and\ \bibinfo {author} {\bibfnamefont {E.~M.}\ \bibnamefont {Lifshitz}},\ }\href {https://archive.org/details/FluidMechanics} {\emph {\bibinfo {title} {Fluid mechanics}}},\ Vol.~\bibinfo {volume} {6}\ (\bibinfo  {publisher} {Pergamon Press, New York},\ \bibinfo {year} {1959})\BibitemShut {NoStop}%
\bibitem [{\citenamefont {Wittkowski}\ \emph {et~al.}(2014)\citenamefont {Wittkowski}, \citenamefont {Tiribocchi}, \citenamefont {Stenhammar}, \citenamefont {Allen}, \citenamefont {Marenduzzo},\ and\ \citenamefont {Cates}}]{wittkowski2014scalar}%
  \BibitemOpen
  \bibfield  {author} {\bibinfo {author} {\bibfnamefont {R.}~\bibnamefont {Wittkowski}}, \bibinfo {author} {\bibfnamefont {A.}~\bibnamefont {Tiribocchi}}, \bibinfo {author} {\bibfnamefont {J.}~\bibnamefont {Stenhammar}}, \bibinfo {author} {\bibfnamefont {R.~J.}\ \bibnamefont {Allen}}, \bibinfo {author} {\bibfnamefont {D.}~\bibnamefont {Marenduzzo}},\ and\ \bibinfo {author} {\bibfnamefont {M.~E.}\ \bibnamefont {Cates}},\ }\bibfield  {title} {\bibinfo {title} {Scalar $\varphi^4$ field theory for active-particle phase separation},\ }\href {https://doi.org/10.1038/ncomms5351} {\bibfield  {journal} {\bibinfo  {journal} {Nat. Comm.}\ }\textbf {\bibinfo {volume} {5}},\ \bibinfo {pages} {4351} (\bibinfo {year} {2014})}\BibitemShut {NoStop}%
\bibitem [{\citenamefont {Doostmohammadi}\ \emph {et~al.}(2018)\citenamefont {Doostmohammadi}, \citenamefont {Ign{\'e}s-Mullol}, \citenamefont {Yeomans},\ and\ \citenamefont {Sagu{\'e}s}}]{doostmohammadi2018active}%
  \BibitemOpen
  \bibfield  {author} {\bibinfo {author} {\bibfnamefont {A.}~\bibnamefont {Doostmohammadi}}, \bibinfo {author} {\bibfnamefont {J.}~\bibnamefont {Ign{\'e}s-Mullol}}, \bibinfo {author} {\bibfnamefont {J.~M.}\ \bibnamefont {Yeomans}},\ and\ \bibinfo {author} {\bibfnamefont {F.}~\bibnamefont {Sagu{\'e}s}},\ }\bibfield  {title} {\bibinfo {title} {Active nematics},\ }\href@noop {} {\bibfield  {journal} {\bibinfo  {journal} {Nat. Commun.}\ }\textbf {\bibinfo {volume} {9}},\ \bibinfo {pages} {3246} (\bibinfo {year} {2018})}\BibitemShut {NoStop}%
\bibitem [{\citenamefont {Alert}\ \emph {et~al.}(2020)\citenamefont {Alert}, \citenamefont {Joanny},\ and\ \citenamefont {Casademunt}}]{alert2020universal}%
  \BibitemOpen
  \bibfield  {author} {\bibinfo {author} {\bibfnamefont {R.}~\bibnamefont {Alert}}, \bibinfo {author} {\bibfnamefont {J.-F.}\ \bibnamefont {Joanny}},\ and\ \bibinfo {author} {\bibfnamefont {J.}~\bibnamefont {Casademunt}},\ }\bibfield  {title} {\bibinfo {title} {Universal scaling of active nematic turbulence},\ }\href@noop {} {\bibfield  {journal} {\bibinfo  {journal} {Nat. Phys.}\ }\textbf {\bibinfo {volume} {16}},\ \bibinfo {pages} {682} (\bibinfo {year} {2020})}\BibitemShut {NoStop}%
\bibitem [{\citenamefont {Mukherjee}\ \emph {et~al.}(2023)\citenamefont {Mukherjee}, \citenamefont {Singh}, \citenamefont {James},\ and\ \citenamefont {Ray}}]{mukherjee2023intermittency}%
  \BibitemOpen
  \bibfield  {author} {\bibinfo {author} {\bibfnamefont {S.}~\bibnamefont {Mukherjee}}, \bibinfo {author} {\bibfnamefont {R.~K.}\ \bibnamefont {Singh}}, \bibinfo {author} {\bibfnamefont {M.}~\bibnamefont {James}},\ and\ \bibinfo {author} {\bibfnamefont {S.~S.}\ \bibnamefont {Ray}},\ }\bibfield  {title} {\bibinfo {title} {Intermittency, fluctuations and maximal chaos in an emergent universal state of active turbulence},\ }\href@noop {} {\bibfield  {journal} {\bibinfo  {journal} {Nat. Phys.}\ }\textbf {\bibinfo {volume} {19}},\ \bibinfo {pages} {891} (\bibinfo {year} {2023})}\BibitemShut {NoStop}%
\bibitem [{\citenamefont {J{\"u}licher}\ \emph {et~al.}(2018)\citenamefont {J{\"u}licher}, \citenamefont {Grill},\ and\ \citenamefont {Salbreux}}]{julicher2018hydrodynamic}%
  \BibitemOpen
  \bibfield  {author} {\bibinfo {author} {\bibfnamefont {F.}~\bibnamefont {J{\"u}licher}}, \bibinfo {author} {\bibfnamefont {S.~W.}\ \bibnamefont {Grill}},\ and\ \bibinfo {author} {\bibfnamefont {G.}~\bibnamefont {Salbreux}},\ }\bibfield  {title} {\bibinfo {title} {Hydrodynamic theory of active matter},\ }\href@noop {} {\bibfield  {journal} {\bibinfo  {journal} {Rep. Prog. Phys.}\ }\textbf {\bibinfo {volume} {81}},\ \bibinfo {pages} {076601} (\bibinfo {year} {2018})}\BibitemShut {NoStop}%
\bibitem [{\citenamefont {Lecun}\ \emph {et~al.}(1998)\citenamefont {Lecun}, \citenamefont {Bottou}, \citenamefont {Bengio},\ and\ \citenamefont {Haffner}}]{lecun1998gradient}%
  \BibitemOpen
  \bibfield  {author} {\bibinfo {author} {\bibfnamefont {Y.}~\bibnamefont {Lecun}}, \bibinfo {author} {\bibfnamefont {L.}~\bibnamefont {Bottou}}, \bibinfo {author} {\bibfnamefont {Y.}~\bibnamefont {Bengio}},\ and\ \bibinfo {author} {\bibfnamefont {P.}~\bibnamefont {Haffner}},\ }\bibfield  {title} {\bibinfo {title} {Gradient-based learning applied to document recognition},\ }\href {https://doi.org/10.1109/5.726791} {\bibfield  {journal} {\bibinfo  {journal} {Proceedings of the IEEE}\ }\textbf {\bibinfo {volume} {86}},\ \bibinfo {pages} {2278} (\bibinfo {year} {1998})}\BibitemShut {NoStop}%
\bibitem [{\citenamefont {Goodfellow}\ \emph {et~al.}(2016)\citenamefont {Goodfellow}, \citenamefont {Bengio},\ and\ \citenamefont {Courville}}]{goodfellow2016deep}%
  \BibitemOpen
  \bibfield  {author} {\bibinfo {author} {\bibfnamefont {I.}~\bibnamefont {Goodfellow}}, \bibinfo {author} {\bibfnamefont {Y.}~\bibnamefont {Bengio}},\ and\ \bibinfo {author} {\bibfnamefont {A.}~\bibnamefont {Courville}},\ }\href@noop {} {\emph {\bibinfo {title} {Deep Learning}}}\ (\bibinfo  {publisher} {MIT Press},\ \bibinfo {year} {2016})\BibitemShut {NoStop}%
\bibitem [{\citenamefont {Hershey}\ \emph {et~al.}(2017)\citenamefont {Hershey}, \citenamefont {Chaudhuri}, \citenamefont {Ellis}, \citenamefont {Gemmeke}, \citenamefont {Jansen}, \citenamefont {Moore}, \citenamefont {Plakal}, \citenamefont {Platt}, \citenamefont {Saurous}, \citenamefont {Seybold}, \citenamefont {Slaney}, \citenamefont {Weiss},\ and\ \citenamefont {Wilson}}]{hershey2017cnn}%
  \BibitemOpen
  \bibfield  {author} {\bibinfo {author} {\bibfnamefont {S.}~\bibnamefont {Hershey}}, \bibinfo {author} {\bibfnamefont {S.}~\bibnamefont {Chaudhuri}}, \bibinfo {author} {\bibfnamefont {D.~P.~W.}\ \bibnamefont {Ellis}}, \bibinfo {author} {\bibfnamefont {J.~F.}\ \bibnamefont {Gemmeke}}, \bibinfo {author} {\bibfnamefont {A.}~\bibnamefont {Jansen}}, \bibinfo {author} {\bibfnamefont {C.}~\bibnamefont {Moore}}, \bibinfo {author} {\bibfnamefont {M.}~\bibnamefont {Plakal}}, \bibinfo {author} {\bibfnamefont {D.}~\bibnamefont {Platt}}, \bibinfo {author} {\bibfnamefont {R.~A.}\ \bibnamefont {Saurous}}, \bibinfo {author} {\bibfnamefont {B.}~\bibnamefont {Seybold}}, \bibinfo {author} {\bibfnamefont {M.}~\bibnamefont {Slaney}}, \bibinfo {author} {\bibfnamefont {R.}~\bibnamefont {Weiss}},\ and\ \bibinfo {author} {\bibfnamefont {K.}~\bibnamefont {Wilson}},\ }\bibfield  {title} {\bibinfo {title} {Cnn architectures for large-scale audio classification},\ }in\ \href {https://arxiv.org/abs/1609.09430} {\emph {\bibinfo
  {booktitle} {International Conference on Acoustics, Speech and Signal Processing (ICASSP)}}}\ (\bibinfo {year} {2017})\BibitemShut {NoStop}%
\bibitem [{\citenamefont {"Maas}\ \emph {et~al.}(2013)\citenamefont {"Maas}, \citenamefont {"Hannun},\ and\ \citenamefont {"Ng}}]{maas2013rectifier}%
  \BibitemOpen
  \bibfield  {author} {\bibinfo {author} {\bibfnamefont {A.~L.}\ \bibnamefont {"Maas}}, \bibinfo {author} {\bibfnamefont {A.~Y.}\ \bibnamefont {"Hannun}},\ and\ \bibinfo {author} {\bibfnamefont {A.~Y.}\ \bibnamefont {"Ng}},\ }\bibfield  {title} {\bibinfo {title} {Rectifier nonlinearities improve neural network acoustic models}\ }(\bibinfo {year} {2013})\BibitemShut {NoStop}%
\bibitem [{\citenamefont {Paszke}\ \emph {et~al.}(2019)\citenamefont {Paszke}, \citenamefont {Gross}, \citenamefont {Massa}, \citenamefont {Lerer}, \citenamefont {Bradbury}, \citenamefont {Chanan}, \citenamefont {Killeen}, \citenamefont {Lin}, \citenamefont {Gimelshein}, \citenamefont {Antiga}, \citenamefont {Desmaison}, \citenamefont {K\"{o}pf}, \citenamefont {Yang}, \citenamefont {DeVito}, \citenamefont {Raison}, \citenamefont {Tejani}, \citenamefont {Chilamkurthy}, \citenamefont {Steiner}, \citenamefont {Fang}, \citenamefont {Bai},\ and\ \citenamefont {Chintala}}]{paszke2019pytorch}%
  \BibitemOpen
  \bibfield  {author} {\bibinfo {author} {\bibfnamefont {A.}~\bibnamefont {Paszke}}, \bibinfo {author} {\bibfnamefont {S.}~\bibnamefont {Gross}}, \bibinfo {author} {\bibfnamefont {F.}~\bibnamefont {Massa}}, \bibinfo {author} {\bibfnamefont {A.}~\bibnamefont {Lerer}}, \bibinfo {author} {\bibfnamefont {J.}~\bibnamefont {Bradbury}}, \bibinfo {author} {\bibfnamefont {G.}~\bibnamefont {Chanan}}, \bibinfo {author} {\bibfnamefont {T.}~\bibnamefont {Killeen}}, \bibinfo {author} {\bibfnamefont {Z.}~\bibnamefont {Lin}}, \bibinfo {author} {\bibfnamefont {N.}~\bibnamefont {Gimelshein}}, \bibinfo {author} {\bibfnamefont {L.}~\bibnamefont {Antiga}}, \bibinfo {author} {\bibfnamefont {A.}~\bibnamefont {Desmaison}}, \bibinfo {author} {\bibfnamefont {A.}~\bibnamefont {K\"{o}pf}}, \bibinfo {author} {\bibfnamefont {E.}~\bibnamefont {Yang}}, \bibinfo {author} {\bibfnamefont {Z.}~\bibnamefont {DeVito}}, \bibinfo {author} {\bibfnamefont {M.}~\bibnamefont {Raison}}, \bibinfo {author} {\bibfnamefont {A.}~\bibnamefont {Tejani}},
  \bibinfo {author} {\bibfnamefont {S.}~\bibnamefont {Chilamkurthy}}, \bibinfo {author} {\bibfnamefont {B.}~\bibnamefont {Steiner}}, \bibinfo {author} {\bibfnamefont {L.}~\bibnamefont {Fang}}, \bibinfo {author} {\bibfnamefont {J.}~\bibnamefont {Bai}},\ and\ \bibinfo {author} {\bibfnamefont {S.}~\bibnamefont {Chintala}},\ }\bibinfo {title} {Pytorch: an imperative style, high-performance deep learning library},\ in\ \href@noop {} {\emph {\bibinfo {booktitle} {Proceedings of the 33rd International Conference on Neural Information Processing Systems}}}\ (\bibinfo  {publisher} {Curran Associates Inc.},\ \bibinfo {address} {Red Hook, NY, USA},\ \bibinfo {year} {2019})\BibitemShut {NoStop}%
\bibitem [{\citenamefont {Loshchilov}\ and\ \citenamefont {Hutter}(2019)}]{loshchilov2019decoupled}%
  \BibitemOpen
  \bibfield  {author} {\bibinfo {author} {\bibfnamefont {I.}~\bibnamefont {Loshchilov}}\ and\ \bibinfo {author} {\bibfnamefont {F.}~\bibnamefont {Hutter}},\ }\bibfield  {title} {\bibinfo {title} {Decoupled weight decay regularization},\ }in\ \href {https://openreview.net/forum?id=Bkg6RiCqY7} {\emph {\bibinfo {booktitle} {International Conference on Learning Representations}}}\ (\bibinfo {year} {2019})\BibitemShut {NoStop}%
\bibitem [{\citenamefont {MacKay}(2003)}]{mackay2003information}%
  \BibitemOpen
  \bibfield  {author} {\bibinfo {author} {\bibfnamefont {D.~J.}\ \bibnamefont {MacKay}},\ }\href@noop {} {\emph {\bibinfo {title} {Information theory, inference and learning algorithms}}}\ (\bibinfo  {publisher} {Cambridge university press},\ \bibinfo {year} {2003})\BibitemShut {NoStop}%
\bibitem [{\citenamefont {Sivia}\ and\ \citenamefont {Skilling}(2006)}]{sivia2006data}%
  \BibitemOpen
  \bibfield  {author} {\bibinfo {author} {\bibfnamefont {D.}~\bibnamefont {Sivia}}\ and\ \bibinfo {author} {\bibfnamefont {J.}~\bibnamefont {Skilling}},\ }\href@noop {} {\emph {\bibinfo {title} {Data analysis: a Bayesian tutorial}}}\ (\bibinfo  {publisher} {OUP Oxford},\ \bibinfo {year} {2006})\BibitemShut {NoStop}%
\bibitem [{\citenamefont {Jaynes}(2003)}]{jaynes2003probability}%
  \BibitemOpen
  \bibfield  {author} {\bibinfo {author} {\bibfnamefont {E.~T.}\ \bibnamefont {Jaynes}},\ }\href@noop {} {\emph {\bibinfo {title} {Probability theory: The logic of science}}}\ (\bibinfo  {publisher} {Cambridge university press},\ \bibinfo {year} {2003})\BibitemShut {NoStop}%
\bibitem [{\citenamefont {Cranmer}\ \emph {et~al.}(2020)\citenamefont {Cranmer}, \citenamefont {Brehmer},\ and\ \citenamefont {Louppe}}]{cranmer2020frontier}%
  \BibitemOpen
  \bibfield  {author} {\bibinfo {author} {\bibfnamefont {K.}~\bibnamefont {Cranmer}}, \bibinfo {author} {\bibfnamefont {J.}~\bibnamefont {Brehmer}},\ and\ \bibinfo {author} {\bibfnamefont {G.}~\bibnamefont {Louppe}},\ }\bibfield  {title} {\bibinfo {title} {The frontier of simulation-based inference},\ }\href {https://doi.org/10.1073/pnas.1912789117} {\bibfield  {journal} {\bibinfo  {journal} {Proc. Natl. Acad. Sci. U.S.A.}\ }\textbf {\bibinfo {volume} {117}},\ \bibinfo {pages} {30055} (\bibinfo {year} {2020})}\BibitemShut {NoStop}%
\bibitem [{\citenamefont {Toni}\ \emph {et~al.}(2008)\citenamefont {Toni}, \citenamefont {Welch}, \citenamefont {Strelkowa}, \citenamefont {Ipsen},\ and\ \citenamefont {Stumpf}}]{toni2008approximate}%
  \BibitemOpen
  \bibfield  {author} {\bibinfo {author} {\bibfnamefont {T.}~\bibnamefont {Toni}}, \bibinfo {author} {\bibfnamefont {D.}~\bibnamefont {Welch}}, \bibinfo {author} {\bibfnamefont {N.}~\bibnamefont {Strelkowa}}, \bibinfo {author} {\bibfnamefont {A.}~\bibnamefont {Ipsen}},\ and\ \bibinfo {author} {\bibfnamefont {M.~P.}\ \bibnamefont {Stumpf}},\ }\bibfield  {title} {\bibinfo {title} {Approximate bayesian computation scheme for parameter inference and model selection in dynamical systems},\ }\href@noop {} {\bibfield  {journal} {\bibinfo  {journal} {J. R. Soc. Interface}\ }\textbf {\bibinfo {volume} {6}},\ \bibinfo {pages} {187} (\bibinfo {year} {2008})}\BibitemShut {NoStop}%
\bibitem [{\citenamefont {{R. Singh}}\ \emph {et~al.}(2018)\citenamefont {{R. Singh}}, \citenamefont {Ghosh},\ and\ \citenamefont {Adhikari}}]{singh2018fast}%
  \BibitemOpen
  \bibfield  {author} {\bibinfo {author} {\bibnamefont {{R. Singh}}}, \bibinfo {author} {\bibfnamefont {D.}~\bibnamefont {Ghosh}},\ and\ \bibinfo {author} {\bibfnamefont {R.}~\bibnamefont {Adhikari}},\ }\bibfield  {title} {\bibinfo {title} {Fast {B}ayesian inference of the multivariate ornstein-uhlenbeck process},\ }\href {https://doi.org/10.1103/PhysRevE.98.012136} {\bibfield  {journal} {\bibinfo  {journal} {Phys. Rev. E}\ }\textbf {\bibinfo {volume} {98}},\ \bibinfo {pages} {012136} (\bibinfo {year} {2018})}\BibitemShut {NoStop}%
\bibitem [{\citenamefont {Sisson}\ \emph {et~al.}(2019)\citenamefont {Sisson}, \citenamefont {Fan},\ and\ \citenamefont {Beaumont}}]{sisson2018handbook}%
  \BibitemOpen
  \bibfield  {author} {\bibinfo {author} {\bibfnamefont {S.}~\bibnamefont {Sisson}}, \bibinfo {author} {\bibfnamefont {Y.}~\bibnamefont {Fan}},\ and\ \bibinfo {author} {\bibfnamefont {M.}~\bibnamefont {Beaumont}},\ }\href {https://books.google.co.in/books?id=5c84zgEACAAJ} {\emph {\bibinfo {title} {Handbook of Approximate Bayesian Computation}}},\ Chapman and Hall/CRC Handbooks of Modern Statistical Methods Series\ (\bibinfo  {publisher} {CRC Press, Taylor and Francis Group},\ \bibinfo {year} {2019})\BibitemShut {NoStop}%
\bibitem [{\citenamefont {Stone}(1974)}]{stone1974val}%
  \BibitemOpen
  \bibfield  {author} {\bibinfo {author} {\bibfnamefont {M.}~\bibnamefont {Stone}},\ }\bibfield  {title} {\bibinfo {title} {Cross-validatory choice and assessment of statistical predictions},\ }\href {https://doi.org/10.1111/j.2517-6161.1974.tb00994.x} {\bibfield  {journal} {\bibinfo  {journal} {Journal of the Royal Statistical Society: Series B (Methodological)}\ }\textbf {\bibinfo {volume} {36}},\ \bibinfo {pages} {111} (\bibinfo {year} {1974})}\BibitemShut {NoStop}%
\bibitem [{\citenamefont {Brier}(1950)}]{brier1950}%
  \BibitemOpen
  \bibfield  {author} {\bibinfo {author} {\bibfnamefont {G.~W.}\ \bibnamefont {Brier}},\ }\bibfield  {title} {\bibinfo {title} {Verification of forecasts expressed in terms of probability},\ }\href {https://doi.org/10.1175/1520-0493(1950)078<0001:VOFEIT>2.0.CO;2} {\bibfield  {journal} {\bibinfo  {journal} {Monthly Weather Review}\ }\textbf {\bibinfo {volume} {78}},\ \bibinfo {pages} {1 } (\bibinfo {year} {1950})}\BibitemShut {NoStop}%
\bibitem [{\citenamefont {Pedregosa}\ \emph {et~al.}(2011)\citenamefont {Pedregosa}, \citenamefont {Varoquaux}, \citenamefont {Gramfort}, \citenamefont {Michel}, \citenamefont {Thirion}, \citenamefont {Grisel}, \citenamefont {Blondel}, \citenamefont {Prettenhofer}, \citenamefont {Weiss}, \citenamefont {Dubourg}, \citenamefont {Vanderplas}, \citenamefont {Passos}, \citenamefont {Cournapeau}, \citenamefont {Brucher}, \citenamefont {Perrot},\ and\ \citenamefont {Duchesnay}}]{scikit-learn}%
  \BibitemOpen
  \bibfield  {author} {\bibinfo {author} {\bibfnamefont {F.}~\bibnamefont {Pedregosa}}, \bibinfo {author} {\bibfnamefont {G.}~\bibnamefont {Varoquaux}}, \bibinfo {author} {\bibfnamefont {A.}~\bibnamefont {Gramfort}}, \bibinfo {author} {\bibfnamefont {V.}~\bibnamefont {Michel}}, \bibinfo {author} {\bibfnamefont {B.}~\bibnamefont {Thirion}}, \bibinfo {author} {\bibfnamefont {O.}~\bibnamefont {Grisel}}, \bibinfo {author} {\bibfnamefont {M.}~\bibnamefont {Blondel}}, \bibinfo {author} {\bibfnamefont {P.}~\bibnamefont {Prettenhofer}}, \bibinfo {author} {\bibfnamefont {R.}~\bibnamefont {Weiss}}, \bibinfo {author} {\bibfnamefont {V.}~\bibnamefont {Dubourg}}, \bibinfo {author} {\bibfnamefont {J.}~\bibnamefont {Vanderplas}}, \bibinfo {author} {\bibfnamefont {A.}~\bibnamefont {Passos}}, \bibinfo {author} {\bibfnamefont {D.}~\bibnamefont {Cournapeau}}, \bibinfo {author} {\bibfnamefont {M.}~\bibnamefont {Brucher}}, \bibinfo {author} {\bibfnamefont {M.}~\bibnamefont {Perrot}},\ and\ \bibinfo {author} {\bibfnamefont
  {E.}~\bibnamefont {Duchesnay}},\ }\bibfield  {title} {\bibinfo {title} {Scikit-learn: Machine learning in {P}ython},\ }\href@noop {} {\bibfield  {journal} {\bibinfo  {journal} {Journal of Machine Learning Research}\ }\textbf {\bibinfo {volume} {12}},\ \bibinfo {pages} {2825} (\bibinfo {year} {2011})}\BibitemShut {NoStop}%
\bibitem [{\citenamefont {Mohapatra}\ \emph {et~al.}(2025)\citenamefont {Mohapatra}, \citenamefont {Kumar}, \citenamefont {Deb}, \citenamefont {Dhomkar},\ and\ \citenamefont {Singh}}]{mohapatra2025inferring}%
  \BibitemOpen
  \bibfield  {author} {\bibinfo {author} {\bibfnamefont {A.}~\bibnamefont {Mohapatra}}, \bibinfo {author} {\bibfnamefont {A.}~\bibnamefont {Kumar}}, \bibinfo {author} {\bibfnamefont {M.}~\bibnamefont {Deb}}, \bibinfo {author} {\bibfnamefont {S.}~\bibnamefont {Dhomkar}},\ and\ \bibinfo {author} {\bibfnamefont {R.}~\bibnamefont {Singh}},\ }\bibfield  {title} {\bibinfo {title} {Inferring activity from the flow field around active colloidal particles using deep learning},\ }\href@noop {} {\bibfield  {journal} {\bibinfo  {journal} {J. Fluid Mech.}\ }\textbf {\bibinfo {volume} {1018}},\ \bibinfo {pages} {R1} (\bibinfo {year} {2025})}\BibitemShut {NoStop}%
\bibitem [{\citenamefont {Bayati}\ and\ \citenamefont {Mallory}(2025)}]{bayati2025inferring}%
  \BibitemOpen
  \bibfield  {author} {\bibinfo {author} {\bibfnamefont {P.}~\bibnamefont {Bayati}}\ and\ \bibinfo {author} {\bibfnamefont {S.~A.}\ \bibnamefont {Mallory}},\ }\bibfield  {title} {\bibinfo {title} {Inferring surface slip in active colloids from flow fields using physics-informed neural networks},\ }\href@noop {} {\bibfield  {journal} {\bibinfo  {journal} {arXiv preprint arXiv:2511.22723}\ } (\bibinfo {year} {2025})}\BibitemShut {NoStop}%
\bibitem [{\citenamefont {Tjhung}\ \emph {et~al.}(2018)\citenamefont {Tjhung}, \citenamefont {Nardini},\ and\ \citenamefont {Cates}}]{tjhung2018cluster}%
  \BibitemOpen
  \bibfield  {author} {\bibinfo {author} {\bibfnamefont {E.}~\bibnamefont {Tjhung}}, \bibinfo {author} {\bibfnamefont {C.}~\bibnamefont {Nardini}},\ and\ \bibinfo {author} {\bibfnamefont {M.~E.}\ \bibnamefont {Cates}},\ }\bibfield  {title} {\bibinfo {title} {Cluster phases and bubbly phase separation in active fluids: Reversal of the ostwald process},\ }\href {https://doi.org/10.1103/PhysRevX.8.031080} {\bibfield  {journal} {\bibinfo  {journal} {Phys. Rev. X}\ }\textbf {\bibinfo {volume} {8}},\ \bibinfo {pages} {031080} (\bibinfo {year} {2018})}\BibitemShut {NoStop}%
\bibitem [{\citenamefont {Pozrikidis}(1992)}]{pozrikidis1992}%
  \BibitemOpen
  \bibfield  {author} {\bibinfo {author} {\bibfnamefont {C.}~\bibnamefont {Pozrikidis}},\ }\href {https://doi.org/10.1017/CBO9780511624124} {\emph {\bibinfo {title} {Boundary Integral and Singularity Methods for Linearized Viscous Flow}}}\ (\bibinfo  {publisher} {Cambridge U.P.},\ \bibinfo {year} {1992})\BibitemShut {NoStop}%
\bibitem [{\citenamefont {Boyd}(2001)}]{boyd2001chebyshev}%
  \BibitemOpen
  \bibfield  {author} {\bibinfo {author} {\bibfnamefont {J.~P.}\ \bibnamefont {Boyd}},\ }\href@noop {} {\emph {\bibinfo {title} {Chebyshev and Fourier spectral methods}}}\ (\bibinfo  {publisher} {Courier Corporation},\ \bibinfo {year} {2001})\BibitemShut {NoStop}%
\bibitem [{\citenamefont {Kutz}(2026)}]{kutz2026data}%
  \BibitemOpen
  \bibfield  {author} {\bibinfo {author} {\bibfnamefont {J.~N.}\ \bibnamefont {Kutz}},\ }\href@noop {} {\emph {\bibinfo {title} {Data-driven modeling \& scientific computation: methods for complex systems \& big data}}}\ (\bibinfo  {publisher} {Oxford University Press},\ \bibinfo {year} {2026})\BibitemShut {NoStop}%
\end{thebibliography}

%%=======================
%%=======================
%%=======================
\end{document}